\documentclass[12pt]{report}
\usepackage{uiuc_thesis}

\usepackage{msthesis}
\usepackage{pstricks, pst-tree}
\usepackage{url}

\mastersthesis

\leftchapter

\oneandhalfspace 

\begin{document}
\renewcommand{\thesistitlepageauthor}{SHRIPAD THITE} 
\renewcommand{\thesisauthor}{Shripad Thite} 

\renewcommand{\previousschools}{B.E., University of Poona, 1997}

\renewcommand{\thesismonth}{September} 
\renewcommand{\thesisyear}{2001}

\renewcommand{\thesistitle}{OPTIMUM BINARY SEARCH TREES ON THE
HIERARCHICAL MEMORY MODEL} 

\renewcommand{\thesisdedication}{To my father}

\renewcommand{\thesissupervisor}{Prof. Michael C. Loui}

\renewcommand{\thesisdepartment}{Computer Science}
\renewcommand{\thesisdegreein}{Computer Science}



\thesistitlepage                     

\thesiscopyrightpage                 


\begin{thesisabstract}               

The Hierarchical Memory Model (HMM) of computation is similar to the
standard Random Access Machine (RAM) model except that the HMM has a
non-uniform memory organized in a hierarchy of levels numbered $1$
through $h$. The cost of accessing a memory location increases with
the level number, and accesses to memory locations belonging to the
same level cost the same. Formally, the cost of a single access to the
memory location at address $a$ is given by $\mu(a)$, where $\mu: \Nat
\rightarrow \Nat$ is the memory cost function, and the $h$ distinct
values of $\mu$ model the different levels of the memory hierarchy.

We study the problem of constructing and storing a binary search tree
(BST) of minimum cost, over a set of keys, with probabilities for
successful and unsuccessful searches, on the HMM with an arbitrary
number of memory levels, and for the special case $h=2$.

While the problem of constructing optimum binary search trees has been
well studied for the standard RAM model, the additional parameter
$\mu$ for the HMM increases the combinatorial complexity of the
problem. We present two dynamic programming algorithms to construct
optimum BSTs bottom-up. These algorithms run efficiently under some
natural assumptions about the memory hierarchy. We also give an
efficient algorithm to construct a BST that is close to optimum, by
modifying a well-known linear-time approximation algorithm for the RAM
model. We conjecture that the problem of constructing an optimum BST
for the HMM with an arbitrary memory cost function $\mu$ is
NP-complete.
\end{thesisabstract}                 

\thesisdedicationpage                

\begin{quote}
``Results? Why, man, I have gotten lots of results! If I find 10,000
ways something won't work, I haven't failed.''\\
\qquad --- Thomas Alva Edison. (\url{www.thomasedison.com})
\end{quote}

\begin{thesisacknowledgments}
First and foremost, I would like to thank my advisor, Michael Loui.
This thesis would have been of much poorer quality if not for the
copious amounts of time and red ink devoted by him. Prof.\ Loui has
been a wonderful and understanding guide and mentor, and I feel
privileged to have had him as an advisor.

Thanks to Jeff Erickson and Sariel Har-Peled for taking the time to
read and suffer early drafts, and for numerous helpful
discussions. Special thanks to Jeff Erickson for letting me spend an
inordinate amount of time on this project while I was supposed to be
working on something else. I am extremely grateful to Mitch Harris for
being there on so many occasions to listen to my ramblings, to bounce
ideas off of, and often just for being there. I would also like to
thank Prof.\ Ed Reingold; it was during his CS 473 class in fall 1998
that the topic of optimum binary search trees (on the RAM model) came
up for discussion.

I would like to thank my mentor at the Los Alamos National Laboratory,
Madhav Marathe, for providing support and an environment in which to
explore the general subject of hierarchical memory models during my
internship there in summer 1998.
\end{thesisacknowledgments}

\tableofcontents                     
\listoffigures


\newpage \pagenumbering{arabic}  

\setcounter{chapter}{0}
\chapter{Introduction}
\label{chap1introduction}

\section{What is a binary search tree?}
For a set of $n$ distinct keys $x_1$, $x_2$, $\ldots$, $x_n$ from a
totally ordered universe ($x_1 \prec x_2 \prec \ldots \prec x_n$), a
binary search tree (BST) $T$ is an ordered, rooted binary tree with
$n$ internal nodes. The internal nodes of the tree correspond to the
keys $x_1$ through $x_n$ such that an inorder traversal of the nodes
visits the keys in order of precedence, i.e., in the order $x_1$, $x_2$,
$\ldots$, $x_n$. The external nodes correspond to intervals between
the keys, i.e., the $j$-th external node represents the set of
elements between $x_{j-1}$ and $x_j$. Without ambiguity, we identify
the nodes of the tree by the corresponding keys.

\begin{figure}[t]
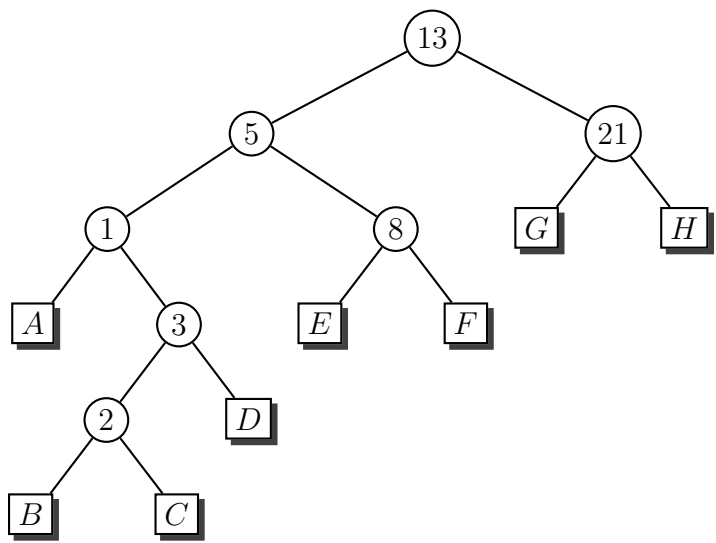

\begin{center}
\pstree[treesep=0.5in, levelsep=0.5in]{\Tcircle{$13$}}{
  \pstree{\Tcircle{$5$}}{
    \pstree{\Tcircle{$1$}}{
      \Trect{$A$}
      \pstree{\Tcircle{$3$}}{
        \pstree{\Tcircle{$2$}}{\Trect{$B$} \Trect{$C$}}
        \Trect{$D$}
      }
    }
    \pstree{\Tcircle{$8$}}{\Trect{$E$} \Trect{$F$}}
  }
  \pstree{\Tcircle{$21$}}{\Trect{$G$} \Trect{$H$}}
}
\end{center}
\caption{A binary search tree over the set $\{1$, $2$, $3$, $5$, $8$, $13$,
$21\}$}
\label{fig_BST}
\hrule
\end{figure}

For instance, a binary search tree on the set of integers $\{1$, $2$,
$3$, $5$, $8$, $13$, $21\}$ with the natural ordering of integers
could look like the tree in figure \ref{fig_BST}. The internal nodes
of the tree are labeled $\{1$, $2$, $3$, $5$, $8$, $13$, $21\}$ and
the external nodes (leaves) are labeled $A$ through $H$ in order.

Let $T_{i,j}$ for $1 \leq i \leq j \leq n$ denote a BST on the
subset of keys from $x_i$ through $x_j$. We define $T_{i+1,i}$ to be
the unique BST over the empty subset of keys from $x_{i+1}$ through
$x_i$ which consists of a single external node with probability of
access $q_i$. We will use $T$ to denote $T_{1,n}$.

A binary search tree with $n$ internal nodes is stored in $n$
locations in memory: each memory location contains a key $x_i$ and two
pointers to the memory locations containing the left and right
children of $x_i$. If the left (resp.\ right) subtree is empty, then
the left (resp.\ right) pointer is \textsc{Nil}.

In this section, we will restrict our attention to the standard RAM
model of computation.

\subsection{Searching in a BST}
A search in $T_{i,j}$ proceeds recursively as follows. The
search argument $y$ is compared with the root $x_k$ ($i \leq k \leq
j$). If $y = x_k$, then the search terminates successfully. Otherwise,
if $y \prec x_k$ (resp.\ $y \succ x_k$), then the search proceeds
recursively in the left subtree, $T_{i,k-1}$ (resp.\ the right subtree,
$T_{k+1,j}$); if the left subtree (resp.\ right subtree) of $x_k$ is an
external node, i.e., a leaf, then the search fails without visiting any
other nodes because $x_{k-1} \prec y \prec x_k$ (resp.\ $x_k \prec y
\prec x_{k+1}$). (We adopt the convention that $x_0 \prec y \prec x_1$
means $y \prec x_1$, and $x_n \prec y \prec x_{n+1}$ means $y \succ
x_n$.)

The depth of an internal or external node $v$ is the number of nodes
on the path to the node from the root, denoted by $\delta_T(v)$, or
simply $\delta(v)$ when the tree $T$ is implicit. Hence, for instance,
the depth of the root is $1$. The cost of a successful or unsuccessful
search is the number of comparisons needed to determine the
outcome. Therefore, the cost of a successful search that terminates at
some internal node $x_i$ is equal to the depth of $x_i$, i.e.,
$\delta(x_i)$.  The cost of an unsuccessful search that would have
terminated at the external node $z_j$ is one less than the depth of
$z_j$, i.e., $\delta(z_j) - 1$.

So, for instance, the depth of the internal node labeled $8$ in the
tree of figure \ref{fig_BST} is $3$. A search for the key $8$ would
perform three comparisons, with the nodes labeled $13$, $5$, and $8$,
before terminating successfully. Therefore, the cost of a successful
search that terminates at the node labeled $8$ is the same as the path
length of the node, i.e., $3$. On the other hand, a search for the
value $4$ would perform comparisons with the nodes labeled $13$, $5$,
$1$, and $3$ in that order and then would terminate with failure, for
a total of four comparisons. This unsuccessful search would have
visited the external node labeled $D$; therefore, the cost of a search
that terminates at $D$ is one less than the depth of $D$, i.e., $5 - 1
= 4$.

Even though the external nodes are conceptually present, they
are not necessary for implementing the BST data structure. If any
subtree of an internal node is empty, then the pointer to that subtree
is assumed to be \NIL; it is not necessary to ``visit'' this empty
subtree.

\subsection{Weighted binary search trees}
In the weighted case, we are also given the probability that the search
argument $y$ is equal to some key $x_i$ for $1 \leq i \leq n$
and the probability that $y$ lies between $x_j$ and $x_{j+1}$ for $0
\leq j \leq n$. Let $p_i$, for $i = 1$, $2$, $\ldots$, $n$, denote the
probability that $y=x_i$. Let $q_j$, for $j = 0$, $1$, $\ldots$, $n$,
denote the probability that $x_j \prec y \prec x_{j+1}$. We have
$$\sum_{i=1}^{n} p_i + \sum_{j=0}^{n} q_j = 1.$$

Define $w_{i,j}$ as
\begin{equation}
\label{eqn_w_ij}
w_{i,j} = \sum_{k=i}^{j} p_k + \sum_{k=i-1}^{j} q_k.
\end{equation}
Therefore, $w_{1,n} = 1$, and $w_{i+1,i} = q_i$. (Note that this
definition differs from the function $w(i,j)$ referred to by Knuth
\cite{KnuthACP3}. Under definition (\ref{eqn_w_ij}), $w_{i,j}$ is the
sum of the probabilities associated with the subtree over the keys
$x_i$ through $x_j$. Under Knuth's definition, $w(i,j) = w_{i+1,j}$ is
the sum of the probabilities associated with the keys $x_{i+1}$
through $x_j$.)

Recall that the cost of a successful search that terminates at the
internal node $x_i$ is $\delta(x_i)$, and the cost of an unsuccessful
search that terminates at the external node $z_j$ is $\delta(z_j) -
1$.  We define the \emph{cost} of $T$ to be the expected cost of a
search:
\begin{equation}
\label{eqn_RAMcost}
\mbox{cost}(T) = \sum_{i=1}^{n} p_i \cdot \delta_T(x_i) +
\sum_{j=0}^{n} q_j \cdot (\delta_T(z_j) - 1).
\end{equation}
In other words, the cost of $T$ is the weighted sum of the depths of
the internal and external nodes of $T$.

An \emph{optimum binary search tree} $T^*$ is one with minimum
cost. Let $T^*_{i,j}$ denote the optimum BST over the subset of keys
from $x_i$ through $x_j$ for all $i$, $j$ such that $1 \leq i \leq j
\leq n$; $T^*_{i+1,i}$ denotes the unique optimum BST consisting of an
external node with probability of access $q_i$.

\section{Why study binary search trees?}
The binary search tree is a fundamental data structure that supports
the operations of inserting and deleting keys, as well as searching
for a key. The straightforward implementation of a BST is adequate and
efficient for the static case when the probabilities of accessing keys
are known \emph{a priori} or can at least be estimated. More
complicated implementations, such as red-black trees \cite{CLR}, AVL
trees \cite{AVLTrees, KnuthACP3}, and splay trees \cite{SplayTrees},
guarantee that a sequence of operations, including insertions and
deletions, can be executed efficiently.

In addition, the binary search tree also serves as a model for
studying the performance of algorithms like \textsc{Quicksort}
\cite{KnuthACP3, CLR}. The recursive execution of \textsc{Quicksort}
corresponds to a binary tree where each node represents a partition of
the elements to be sorted into left and right parts, consisting of
elements that are respectively less than and greater than the pivot
element. The running time of \textsc{Quicksort} is the sum of the work
done by the algorithm corresponding to each node of this recursion
tree.

A binary search tree also arises implicitly in the context of binary
search. The BST corresponding to binary search achieves the
theoretical minimum number of comparisons that are necessary to search
using only key comparisons.

When an explicit BST is used as a data structure, we want to construct
one with minimum cost. When studying the performance of
\textsc{Quicksort}, we want to prove lower bounds on the cost and
hence the running time. Therefore, the problem of constructing optimum
BSTs is of considerable interest.

\section{Overview}
In chapter \ref{chap2background}, we survey background work on binary
search trees and computational models for non-uniform memory
computers.

In chapter \ref{chap3main}, we give algorithms for constructing
optimum binary search trees. In section \ref{secBST4HMM}, we consider
the most general variant of the HMM model, with an arbitrary number of
memory levels. We present two dynamic programming algorithms and a
top-down algorithm to construct optimum BSTs on the HMM. In section
\ref{secBST4HMM2}, we consider the special case of the HMM model with
only two memory levels. For this model, we present a dynamic
programming algorithm to construct optimum BSTs in section
\ref{secdynprog4hmm2}, and in section \ref{secApproxBST}, a
linear-time heuristic to construct a BST close to the optimum.

Finally, we conclude with a summary of our results and a discussion of open
problems in chapter \ref{chap4openconclusions}.


\setcounter{chapter}{1}
\chapter{Background and Related Work}
\label{chap2background}

In this chapter, we survey related work on the problem of constructing
optimum binary search trees, and on computational models for
hierarchical memory. In section \ref{secBackground} we discuss the
optimum binary search tree problem and related problems. In section
\ref{secComputationalModels}, we discuss memory effects in modern
computers and present arguments for better theoretical models. In
section \ref{subsecExternalMemAlg}, we survey related work on
designing data structures and algorithms, and in section
\ref{subsecModels}, we discuss proposed models of computation for
hierarchical-memory computers.

\section{Binary search trees and related problems}
\label{secBackground}
The binary search tree has been studied extensively in different
contexts. In sections \ref{secOptBST4RAM} through \ref{secOptimalBDT},
we will summarize previous work on the following related problems that
have been studied on the RAM model of computation:
\begin{itemize}
\item constructing a binary search tree such that the expected cost of
a search is minimized;
\item constructing an alphabetic tree such that the sum of the
weighted path lengths of the external nodes is minimized;
\item constructing a prefix-free code tree with no restriction on the
lexicographic order of the nodes such that the weighted path lengths
of all nodes is minimized;
\item constructing a binary search tree close to optimum by an
efficient heuristic;
\item constructing an optimal binary decision tree.
\end{itemize}

\subsection{Constructing optimum binary search trees on the RAM}
\label{secOptBST4RAM}

\subsubsection{Dynamic programming algorithms}
\begin{theorem}[Knuth \cite{KnuthBST}, \cite{KnuthACP3}]
\label{thm_n2algorithm}
An optimum BST can be constructed by a dynamic programming algorithm
that runs in $O(n^2)$-time and $O(n^2)$-space.
\end{theorem}
\begin{proof}
By the principle of optimality, a binary search tree $T^*$ is
optimum if and only if each subtree of $T^*$ is optimum. The standard dynamic
programming algorithm proceeds as follows:

Recall that $cost(T^*_{i,j})$ denotes the cost of an optimum BST
$T^*_{i,j}$ over the keys $x_i$, $x_{i+1}$, $\ldots$, $x_j$
and the corresponding probabilities $p_i$, $p_{i+1}$, $\ldots$, $p_j$
and $q_{i-1}$, $q_i$, $\ldots$, $q_j$. By the principle of optimality
and the definition of the cost function in equation (\ref{eqn_RAMcost}),
\begin{align}
\label{eqn_recurrenceBST}
\text{cost}(T^*_{i,j}) &= w_{i,j} + \min_{i \leq k \leq
j}\Paren{\text{cost}(T^*_{i,k-1}) + \text{cost}(T^*_{k+1,j})}
&\text{for $i \leq j$}
\nonumber\\
\text{cost}(T^*_{i+1,i}) &= w_{i+1,i} = q_i
\end{align}

Recurrence (\ref{eqn_recurrenceBST}) suggests a dynamic programming
algorithm, \textsc{algorithm K1} in figure \ref{fig_algorithmK1}, that
constructs optimum subtrees bottom-up. \textsc{algorithm K1} is the
standard dynamic programming algorithm. For each $d$ from $0$ through
$n-1$, and for each $i$, $j$ such that $j-i=d$, the algorithm
evaluates the cost of a subtree with $x_k$ as the root, for every
possible choice of $k$ between $i$ and $j$, and selects the one for
which this cost is minimized.

\textsc{algorithm K1} constructs arrays $c$ and $r$, such that $c[i,j]$ is the
cost of an optimum BST $T^*_{i,j}$ over the subset of keys from $x_i$
through $x_j$ and $r[i,j]$ is the index of the root of such an optimum
BST. The structure of the tree can be retrieved in $O(n)$ time from
the array $r$ at the end of the algorithm as follows. Let $T[i,j]$
denote the optimum subtree constructed by \textsc{algorithm K1} and
represented implicitly using the array $r$. The index of the root of
this subtree is given by the array entry $r[i,j]$. Recursively, the
left and right subtrees of the root are $T[i, r[i,j]-1]$ and
$T[r[i,j]+1, j]$ respectively.

For each fixed $d$ and $i$, the algorithm takes $O(d)$ time to
evaluate the choice of $x_k$ as the root for all $k$ such that $i \leq
k \leq j=i+d$, and hence, $\sum_{d=0}^{n-1} \sum_{i=1}^{n-d} O(d) =
O(n^3)$ time overall.

\begin{figure}
\begin{center}\fbox{
\begin{algorithm}
\label{n3algorithm}
\textul{\textsc{algorithm K1}$([p_1..p_n],[q_0..q_n])$:}
\\
\\  \comment{Initialization phase.}
\\  \comment{An optimum BST over the empty subset of keys from
$x_{i+1}$ through $x_i$}
\\  \comment{consists of just the external node with probability
$q_i$.}
\\  \comment{The root of this subtree is undefined.}
\\  for $i$ := $0$ to $n$ \+
\\    $c[i+1,i] \leftarrow w_{i+1,i} = q_i$
\\    $r[i+1,i] \leftarrow \NIL$ \-
\\
\\  for $d$ := $0$ to $n-1$ \+
\\    for $i$ := $1$ to $n-d$ \+
\\      $j \leftarrow i + d$
\\
\\      \comment{Initially, the optimum subtree $T^*_{i,j}$ is unknown.}
\\      $c[i,j] \leftarrow \infty$
\\
\\      for $k$ := $i$ to $j$ \+
\\        Let $T'$ be the tree with $x_k$ at the root, and $T^*_{i,k-1}$
and $T^*_{k+1,j}$ \+
\\          as the left and right subtrees, respectively, i.e.,
\\          \pstree[treesep=0.5in,levelsep=0.5in]
            {\Tcircle{$x_k$}}{\Trect{$T^*[i,k-1]$} \Trect{$T^*[k+1,j]$}} \-
\\
\\        Let $c'$ be the cost of $T'$: \+
\\        $c' \leftarrow w_{i,j} + c[i,k-1] + c[k+1,j]$ \-
\\        \comment{Is $T'$ better than the minimum-cost tree so far?}
\\        if $c' < c[i,j]$ \+
\\          $r[i,j] \leftarrow k$
\\          $c[i,j] \leftarrow c'$ \-\-\-\-
\end{algorithm}
}\end{center}
\caption{\textsc{algorithm K1}}
\label{fig_algorithmK1}
\hrule
\end{figure}

Knuth \cite{KnuthBST} showed that the following monotonicity principle
can be used to reduce the time complexity to $O(n^2)$: for all $i$,
$j$, $1 \leq i \leq j \leq n$, let $R(i,j)$ denote the index of the
root of an optimum BST over the keys $x_i$, $x_{i+1}$, $\ldots$,
$x_j$ (if more than one root is optimum, let $R(i,j)$ be the smallest
such index); then
\begin{equation}
\label{eqn_monotonicity}
R(i,j-1) \leq R(i,j) \leq R(i+1,j).
\end{equation}

Therefore, the innermost loop in \textsc{algorithm K1} can be modified
to produce \textsc{algorithm K2} (figure \ref{fig_algorithmK2}) with
improved running time.

\begin{figure}
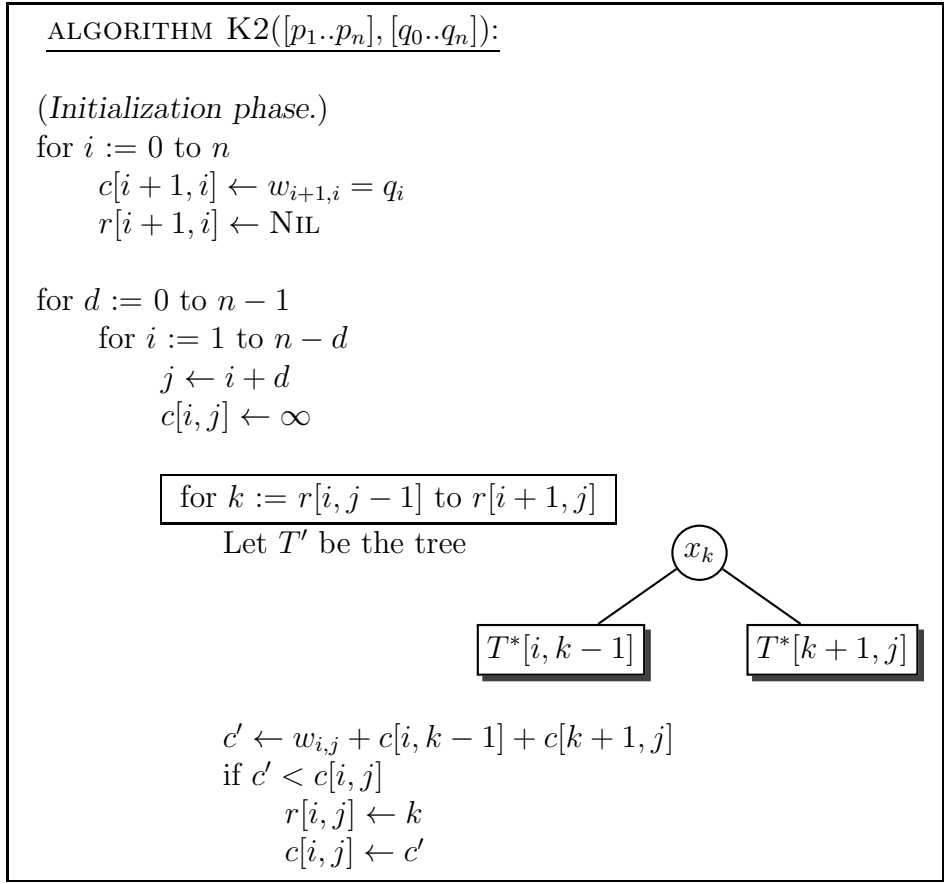

\begin{center}\fbox{
\begin{algorithm}
\label{n2algorithm}
\textul{\textsc{algorithm K2}$([p_1..p_n],[q_0..q_n])$:}
\\
\\  \comment{Initialization phase.}
\\  for $i$ := $0$ to $n$ \+
\\    $c[i+1,i] \leftarrow w_{i+1,i} = q_i$
\\    $r[i+1,i] \leftarrow \NIL$ \-
\\
\\  for $d$ := $0$ to $n-1$ \+
\\    for $i$ := $1$ to $n-d$ \+
\\      $j \leftarrow i + d$
\\      $c[i,j] \leftarrow \infty$
\\
\\      \fbox{ for $k$ := $r[i,j-1]$ to $r[i+1,j]$ } \+
\\        Let $T'$ be the tree \pstree[treesep=0.5in,levelsep=0.5in]
          {\Tcircle{$x_k$}}{\Trect{$T^*[i,k-1]$} \Trect{$T^*[k+1,j]$}}
\\
\\        $c' \leftarrow w_{i,j} + c[i,k-1] + c[k+1,j]$
\\        if $c' < c[i,j]$ \+
\\          $r[i,j] \leftarrow k$
\\          $c[i,j] \leftarrow c'$ \-\-\-\-
\end{algorithm}
}\end{center}
\caption{\textsc{algorithm K2}}
\label{fig_algorithmK2}
\hrule
\end{figure}

Since $(j-1)-i = j-(i+1) = d-1$ whenever $j-i=d$, the values of
$r[i,j-1]$ and $r[i+1,j]$ are available during the iteration when
$j-i=d$. The number of times that the body of the innermost loop in
\textsc{algorithm K2} is executed is $r[i+1,j] - r[i,j-1] + 1$ when
$j-i = d$. Therefore, the running time of \textsc{algorithm K2} is
proportional to
\begin{align*}
&\sum_{d=0}^{n-1} \sum_{i=1}^{n-d} \Paren{r[i+1,j] - r[i,j-1] + 1} \\
  &\qquad\text{where $j = i+d$} \\
&= \sum_{d=0}^{n-1} \Paren{r[n-d+1,n+1] - r[1,d] + n-d} \\
& \leq \sum_{d=0}^{n-1} \Paren{2n-d} \\
  &\qquad\text{since $r[n-d+1,n+1] - r[1,d] \leq (n+1)-1$} \\
& = O(n^2).
\end{align*}

The use of the monotonicity principle above is in fact an application
of the general technique due to Yao \cite{Yao} to speed-up dynamic
programming under some special conditions. (See subsection
\ref{secReproduceYao} below.)

The space required by both algorithms for the tables $r$ and $c$ is
$O(n^2)$.
\end{proof}

\subsubsection{Speed-up in dynamic programming}
\label{secReproduceYao}
For the sake of completeness, we reproduce below results due to Yao
\cite{Yao}.

Consider a recurrence to compute the value of $c(1,n)$ for the
function $c()$ defined by the following recurrence
\begin{align}
\label{eqn_recurrence}
c(i,j) &= w(i,j) + \min_{i \leq k \leq j}\ \Paren{ c(i,k-1) + c(k+1, j)
} &\text{for\ } 1 \leq i \leq j \leq n \nonumber\\
c(i+1,i) &= q_i
\end{align}
where $w()$ is some function and $q_i$ is a constant for $1 \leq i
\leq n$. The form of the recurrence suggests a simple dynamic
programming algorithm that computes $c(i,j)$ from $c(i,k-1)$ and
$c(k+1,j)$ for all $k$ from $i$ through $j$. This algorithm spends
$O(j-i)$ time computing the optimum value of $c(i,j)$ for every pair
$i$, $j$, such that $1 \leq i \leq j \leq n$, for a total running time
of $\sum_{i=1}^{n} \sum_{j=i}^{n} O(j-i) = O(n^3)$.

The function $w(i,j)$ satisfies the \emph{concave quadrangle
inequality} (QI) if:
\begin{equation}
\label{eqn_concaveQI}
w(i,j) + w(i',j') \leq w(i',j) + w(i,j')
\end{equation}
for all $i$, $i'$, $j$, $j'$ such that $i \leq i' \leq j \leq j'$.
In addition, $w(i,j)$ is \emph{monotone} with respect to set inclusion of
intervals if $w(i,j) \leq w(k,l)$ whenever $[i,j] \subseteq [k,l]$,
i.e., $k \leq i \leq j \leq l$.

Let $c_k(i,j)$ denote $w(i,j) + c(i,k-1) + c(k+1,j)$ for each $k$, $i
\leq k \leq j$. Let $K(i,j)$ denote the maximum $k$ for which the
optimum value of $c(i,j)$ is achieved in recurrence 
(\ref{eqn_recurrence}), i.e., for $i \leq j$,
\[
K(i,j) = \max\set{k \mid c_k(i,j) = c(i,j)}
\]
Hence, $K(i,i) = i$.

\begin{lemma}[Yao \cite{Yao}]
If $w(i,j)$ is monotone and satisfies the concave quadrangle
inequality (\ref{eqn_concaveQI}), then the function $c(i,j)$ defined
by recurrence (\ref{eqn_recurrence}) also satisfies the concave QI,
i.e.,
\[
c(i,j) + c(i',j') \leq c(i',j) + c(i,j')
\]
for all $i$, $i'$, $j$, $j'$ such that $i \leq i' \leq j \leq j'$.
\end{lemma}
\begin{proof}[Mehlhorn \cite{Mehlhorn}]
Consider $i$, $i'$, $j$, $j'$ such that $1 \leq i \leq i' \leq j \leq
j' \leq n$. The proof of the lemma is by induction on $l = j' - i$. 

\noindent\textbf{Base cases:} 
The case $l=0$ is trivial.
If $l=1$, then either $i = i'$ or $j = j'$, so the inequality
$$c(i,j) + c(i',j') \leq c(i',j) + c(i,j')$$
is trivially true.

\noindent\textbf{Inductive step:}
Consider the two cases: $i'=j$ and $i' < j$.

\noindent\textbf{Case 1: $i'=j$.} In this case, the concave QI reduces to the
inequality:
$$c(i,j) + c(j,j') \leq c(i,j') + w(j,j).$$

Let $k = K(i,j')$. Clearly, $i \leq k \leq j'$.

\noindent\textbf{Case 1a: $k+1 \leq j$.}
\begin{align*}
c(i,j) + c(j,j') &\leq w(i,j) + c(i,k-1) + c(k+1,j) + c(j,j') \\
&\qquad\text{by the definition of $c(i,j)$} \\
 &\leq w(i,j') + c(i,k-1) + c(k+1,j) + c(j,j') \\
 &\qquad\text{by the monotonicity of $w()$}
\end{align*}
Now if $k+1 \leq j$, then from the induction hypothesis, $c(k+1,j) +
c(j,j') \leq c(k+1,j') + w(j,j)$. Therefore,
\begin{align*}
c(i,j) + c(j,j') &\leq w(i,j') + c(i,k-1) + c(k+1,j') + w(j,j) \\
  &= c(i,j') + w(j,j) \\
  &\qquad\text{because $k=K(i,j')$, and by definition of $c(i,j')$.}
\end{align*}

\noindent\textbf{Case 1b: $k \geq j$.}
\begin{align*}
c(i,j) + c(j,j') &\leq c(i,j) + w(j,j') + c(j,k-1) + c(k+1,j') \\
 &\qquad\text{by the definition of $c(j,j')$} \\
 &\leq c(i,j) + w(i,j') + c(j,k-1) + c(k+1,j') \\
 &\qquad\text{by the monotonicity of $w()$}
\end{align*}
Now if $k \geq j$, then from the induction hypothesis, $c(i,j) +
c(j,k-1) \leq c(i,k-1) + w(j,j)$. Therefore,
\begin{align*}
c(i,j) + c(j,j') &\leq w(i,j') + c(i,k-1) + w(j,j) + c(k+1,j') \\
 &= c(i,j') + w(j,j) \\
 &\qquad\text{by the definition of $c(i,j')$.}
\end{align*}

\noindent\textbf{Case 2: $i' < j$.} Let $y = K(i',j)$ and $z =
K(i,j')$.

\noindent\textbf{Case 2a: $z \leq y$.} Note that $i \leq z \leq y \leq j$.
\begin{align*}
c(i',j') + c(i,j) &= c_y(i',j') + c_z(i,j) \\
 &= \Paren{w(i',j') + c(i',y-1) + c(y+1,j')} + \Paren{w(i,j) +
c(i,z-1) + c(z+1,j)} \\
 &\leq \Paren{w(i,j') + w(i',j')} + \Paren{c(i',y-1) + c(i,z-1) +
c(z+1,j) + c(y+1,j')} \\
 &\qquad\text{from the concave QI for\ } w \\
 &\leq \Paren{w(i,j') + w(i',j')} + \Paren{c(i',y-1) + c(i,z-1) +
c(y+1,j) + c(z+1,j')} \\
 &\qquad\text{from the induction hypothesis,} \\
 &\qquad\text{i.e., the concave QI applied to $z \leq y \leq j \leq j'$} \\
 &= c(i,j') + c(i',j) \\
 &\qquad\text{by definition of $c(i,j')$ and $c(i',j)$.}
\end{align*}

\noindent\textbf{Case 2b: $y \leq z$.} This case is symmetric to case
2a above.
\end{proof}

\begin{theorem}[Yao \cite{Yao}]
\label{thm_Yao}
If the function $w(i,j)$ is monotone and satisfies the concave
quadrangle inequality, then
\[
  K(i,j-1) \leq K(i,j) \leq K(i+1,j).
\]
\end{theorem}
\begin{proof}[Mehlhorn \cite{Mehlhorn}]
The theorem is trivially true when $j=i+1$ because $i \leq K(i,j) \leq
j$. We will prove $K(i,j-1) \leq K(i,j)$ for the case $i<j-1$, by
induction on $j-i$.

Recall that $K(i,j-1)$ is the largest
index $k$ that achieves the minimum value of $c(i,j-1) = w(i,j-1) +
c(i,k-1) + c(k+1,j-1)$ (cf.\ equation (\ref{eqn_recurrence})).
Therefore, it suffices to show that 
\[
c_{k'}(i,j-1) \leq c_k(i,j-1) \implies c_{k'}(i,j) \leq c_k(i,j)
\]
for all $i \leq k \leq k' \leq j$. We prove the stronger inequality
\[
c_k(i,j-1) - c_{k'}(i,j-1) \leq c_k(i,j) - c_{k'}(i,j)
\]
which is equivalent to
\[
c_k(i,j-1) + c_{k'}(i,j) \leq c_{k'}(i,j-1) + c_k(i,j).
\]
The last inequality above is expanded to
\begin{align*}
& c(i,k-1) + c(k+1,j-1) + c(i,k'-1) + c(k'+1,j) \\
&\leq c(i,k'-1) + c(k'+1,j-1) + c(i,k-1) + c(k+1,j)
\end{align*}
or
\[
c(k+1,j-1) + c(k'+1,j) \leq c(k'+1,j-1) + c(k+1,j).
\]
But this is simply the concave quadrangle inequality for the function
$c(i,j)$ for $k \leq k' \leq j-1 \leq j$, which is true by the
induction hypothesis.
\end{proof}


As a consequence of theorem \ref{thm_Yao}, if we compute $c(i,j)$ by
diagonals, in order of increasing values of $j-i$, then we can limit
our search for the optimum value of $k$ to the range from $K(i,j-1)$
through $K(i-1,j)$. The cost of computing all entries on one diagonal
where $j = i+d$ is
\begin{align*}
& \sum_{i=1}^{n-d} \Paren{K(i+1,j) - K(i,j-1) + 1} \\
&= K(n-d+1,n+1) - K(1,d) + n-d \\
&\leq (n+1) - 1 + (n-d) \\
&< 2n.
\end{align*}

The speed-up technique in this section is used to improve the
running time of the standard dynamic programming algorithm to compute
optimum BSTs. It is easy to see that the parameters of the optimum BST
problem satisfy the conditions required by Theorem \ref{thm_Yao}.

\subsection{Alphabetic trees}
\label{secAlphabeticTrees}
The special case of the problem of constructing an optimum BST when
$p_1 = p_2 = \cdots = p_n = 0$ is known as the \emph{alphabetic tree
problem}. This problem arises in the context of constructing optimum
binary code trees. A binary codeword is a string of $0$'s and $1$'s. A
prefix-free binary code is a sequence of binary codewords such that no
codeword is a prefix of another. Corresponding to a prefix-free code
with $n+1$ codewords, there is a rooted binary tree with $n$ internal
nodes and $n+1$ external nodes where the codewords correspond to the
external nodes of the tree.

In the alphabetic tree problem, we require that the codewords at the
external nodes appear in order from left to right. Taking the left
branch of the tree stands for a $0$ bit and taking the right branch
stands for a $1$ bit in the codeword; thus, a path in the tree from
the root to the $j$-th external node represents the bits in
the $j$-th codeword. This method of coding preserves the lexicographic
order of messages. The probability $q_j$ of the $j$-th codeword
is the likelihood that the symbol corresponding to that codeword will
appear in any message. Thus, in this problem, $p_1 = p_2 = \cdots =
p_n = 0$ and $\sum_{j=0}^{n} q_j = 1$.

Hu and Tucker \cite{HuTucker} developed a two-phase algorithm that
constructs an optimum alphabetic tree. In the first phase, starting
with a sequence of $n+1$ nodes, pairs of nodes are recursively
combined into a single tree to obtain an assignment of level numbers
to the nodes. The tree constructed in the first phase does not
necessarily have the leaves in order. In the second phase, the nodes
are recombined into a tree where the nodes are now in lexicographic
order and the depth of a node is the same as the level number assigned
to it in the first phase. It is non-trivial to prove that there exists
an optimum alphabetic tree with the external nodes at the
same depths as the level numbers constructed in the first phase.

The algorithm uses a priority queue with at most $n+1$ elements on
which it performs $O(n)$ operations. With the appropriate
implementation, such as a leftist tree \cite{KnuthACP3} or a Fibonacci
heap \cite{CLR}, the algorithm requires $O(n \log n)$ time and $O(n)$
space.

\subsection{Huffman trees}
\label{secHuffmanTrees}
If we relax the condition in the alphabetic tree problem that the
codewords should be in lexicographic order, then the problem of
constructing an optimum prefix-free code is the Huffman tree
problem. Huffman's classic result
\cite{Huffman} is that a simple greedy algorithm, running in time $O(n
\log n)$, suffices to construct a minimum-cost code tree.

\subsection{Nearly optimum search trees}
\label{secApproximation}
The best known algorithm, \textsc{algorithm K2} due to Knuth
\cite{KnuthBST}, to construct an optimum search tree requires $O(n^2)$
time and space (Theorem \ref{thm_n2algorithm}). If we are willing to
sacrifice optimality for efficiency, then we can use a simple
linear-time heuristic due to Mehlhorn \cite{Mehlhorn} to construct a
tree $T$ that is not too far from optimum. In fact, if $T^*$ is a tree
with minimum cost, then
\[
\text{cost}(T) - \text{cost}(T^*) \leq \lg\Paren{\text{cost}(T^*)}
\approx \lg H
\]
where $H = \sum_{i=1}^{n} p_i \lg (1/p_i) + \sum_{j=0}^{n} q_j \lg
(1/q_j)$ is the entropy of the probability distribution.

\subsection{Optimal binary decision trees}
\label{secOptimalBDT}
We remark that the related problem of constructing an optimal binary
decision tree is known to be NP-complete. Hyafil and Rivest \cite{NPcomplete}
proved that the following problem is NP-hard:
\begin{problem}
Let $S = \{s_1$, $s_2$, $\ldots$, $s_n\}$ be a finite set of objects
and let $T = \{t_1$, $t_2$, $\ldots$, $t_m\}$ be a finite set of
tests. For each test $t_i$ and object $x_j$, $1 \leq i \leq m$ and $1
\leq j \leq n$, we have either $t_i(x_j) = \TRUE$ or $t_i(x_j) =
\FALSE$. Construct an identification procedure for the objects in $S$
such that the expected number of tests required to completely identify
an element of $S$ is minimal. In other words, construct a binary
decision tree with the tests at the internal nodes and the objects in
$S$ at the external nodes, such that the sum of the path lengths of
the external nodes is minimized.
\end{problem}

The authors showed, via a reduction from \textsc{Exact Cover by 3-Sets}
(X3C) \cite{GareyJohnson}, that the optimal binary decision tree problem
remains NP-hard even when the tests are all subsets of $S$ of size
$3$ and $t_i(x_j) = \TRUE$ if and only if $x_j$ is an element of set
$t_i$.

For more details on the optimum binary search tree problem and related
problems, we refer the reader to the excellent survey article by
S.~V. Nagaraj \cite{survey}.

\section{Models of computation}
\label{secComputationalModels}
The Random Access Machine (RAM) \cite{Papadimitriou, BovetCrescenzi}
is used most often in the design and analysis of algorithms.

\subsection{The need for an alternative to the RAM model}
The RAM is a sequential model of computation. It consists of a single
processor with a predetermined set of instructions. Different variants
of the RAM model assume different instruction sets---for instance,
the real RAM \cite{PreparataShamos} can perform exact arithmetic on
real numbers. See also Louis Mak's Ph.D. thesis \cite{Mak}.

In the RAM model, memory is organized as a potentially unbounded array
of locations, numbered $1$, $2$, $3$, $\ldots$, each of which can
store an arbitrarily large integer value. On the RAM, the memory
organization is uniform; i.e., it takes the same amount of time to
access any location in memory.

While the RAM model serves to approximate a real computer fairly well,
in some cases, it has been observed empirically that algorithms (and
data structures) behave much worse than predicted on the RAM
model: their running times are substantially larger than what even a
careful analysis on the RAM model would predict because of memory
effects such as paging and caching. In the following subsections, we
review the hierarchical memory organization of modern computers, and
how it leads to memory effects so that the cost of accessing memory
becomes a significant part of the total running time of an
algorithm. We survey empirical observations of these memory effects,
and the study of data structures and algorithms that attempt to
overcome bottlenecks due to slow memory.

\subsubsection{Modern computer organization}
Modern computers have a hierarchical memory organization
\cite{HennessyPatterson}. Memory is organized into levels such as the
processor's registers, the cache (primary and secondary), main memory,
secondary storage, and even distributed memory. 

The first few levels of the memory hierarchy comprising the CPU
registers, cache, and main memory are realized in silicon components,
i.e., hardware devices such as integrated circuits. This
type of fast memory is called ``internal'' storage, while the slower
magnetic disks, CD-ROMs, and tapes used for realizing secondary and
tertiary storage comprise the ``external'' storage.

Registers have the smallest access time, and magnetic disks and tapes
are the slowest. Typically, the memory in one level is an order of
magnitude faster than in the next level. So, for instance, access
times for registers and cache memory are a few nanoseconds, while
accessing main memory takes tens of nanoseconds.

The sizes (numbers of memory locations) of the levels also increase by
an order of magnitude from one level to the next. So, for instance,
typical cache sizes are measured in kilobytes while main memory sizes
are of the order of megabytes and larger. The reason for these
differences is that faster memory is more expensive to manufacture and
therefore is available in smaller quantities.

Most multi-programmed systems allow the simultaneous execution of
programs in a time-sharing fashion even when the sum of the memory
requirements of the programs exceeds the amount of physical main
memory available. Such systems implement \emph{virtual memory}: not
all data items referenced by a program need to reside in main
memory. The virtual address space, which is much larger than the real
address space, is usually partitioned into pages. Pages can reside
either in main memory or on disk. When the processor references an
address belonging to a page not currently in the main memory, the page
must be loaded from disk into main memory. Therefore, the time to
access a memory location also depends on whether the corresponding
page of virtual memory is currently in main memory.

Consequently, the memory organization is highly non-uniform, and the
assumption of uniform memory cost in the RAM model is unrealistic.

\subsubsection{Locality of reference}
Many algorithms exhibit the phenomenon of spatial and temporal
locality \cite{AJSmith}. Data items are accessed in
regular patterns so that the next item to be accessed is very likely
to be one that is stored close to the last few items accessed. This
phenomenon is called spatial locality. It occurs because data items
that are logically ``close'' to each other also tend to be stored
close together in memory. For instance, an array is a typical data
structure used to represent a list of related items of the same
type. Consecutive array elements are also stored in adjacent memory
locations. (See, however, Chatterjee et al.\
\cite{NonLinearArrayLayout} for a study of the advantage of a
nonlinear layout of arrays in memory. Also, architectures with
interleaved memory store consecutive array elements on different
memory devices to facilitate parallel or pipelined access to a block
of addresses.)

A data item that is accessed at any time is likely to be accessed
again in the near future. For example, the index variable in a loop is
probably also used in the body of the loop. Therefore, during the
execution of the loop, the variable is accessed several times in quick
succession. This is the phenomenon of temporal locality.

In addition, the hardware architecture mandates that the processor can
operate only on data present in its registers. Therefore, executing an
operation requires extra time to move the operands into registers and
store the result back to free up the registers for the next
operation. Typically, data can be moved only between adjacent levels
in the memory hierarchy, such as between the registers and the primary
cache, cache and main memory, and the main memory and secondary
storage, but not directly between the registers and secondary storage.

Therefore, an algorithm designer must make efficient use of available
memory, so that data is available in the fastest possible memory
location whenever it is required. Of course, moving data around
involves extra overhead.  The memory allocation problem is complicated
by the dynamic nature of many algorithms.

\subsubsection{Memory effects}
The effects of caches on the performance of algorithms have been
observed in a number of contexts. Smith \cite{AJSmith} presented a
large number of empirical results obtained by simulating the data
access patterns of real programs on different cache architectures.
LaMarca and Ladner \cite{CacheSorting} investigated the effect of
caches on the performance of sorting algorithms, both experimentally
and analytically. The authors showed how to restructure \textsc{MergeSort},
\textsc{QuickSort}, and \textsc{HeapSort} to improve the utilization
of the cache and reduce the execution time of these
algorithms. Their theoretical prediction of cache misses incurred
closely matches the empirically observed performance.

LaMarca and Ladner \cite{CachedHeaps} also investigated empirically
the performance of heap implementations on different
architectures. They presented optimizations to reduce the cache misses
incurred by heaps and gave empirical data about how their
optimizations affected overall performance on a number of different
architectures.

The performance of several algorithms such as matrix transpositions
and FFT on the virtual memory model was studied by Aggarwal and
Chandra \cite{VirtualMemAlg}. The authors modeled virtual memory as a
large flat address-space which is partitioned into blocks. Each block
of virtual memory is mapped into a block of real memory. A block of
memory must be loaded into real memory before it can be
accessed. The authors showed that some algorithms must still run slowly
even if the algorithms were able to predict memory accesses in advance.

\subsubsection{Complexity of communication}
Algorithms that operate on large data sets spend a substantial amount
of time accessing data (reading from and writing to
memory). Consequently, memory access time (also referred to in the
literature as I/O- or communication-time) frequently dominates the
computation time. Therefore, the RAM model, which does not account for
memory effects, is inadequate for accurately predicting the
performance of such algorithms.

Depending on the machine organization, either the time to compute
results or the time to read/write data may dominate the running time
of an algorithm. A computation graph represents the dependency relationship
between data items---there is a directed edge from vertex $u$ to
vertex $v$ if the operation that computes the value at $v$ requires
that the value at $u$ be already available. For computation on a
collection of values whose dependencies form a grid graph, the
tradeoff between the computation time and memory access time was
quantified by Papadimitriou and Ullman \cite{CommunicationVsTime}.

The I/O-complexity of an algorithm is the cost of inputs and outputs
between faster internal memory and slower secondary 
memory. Aggarwal and Vitter \cite{IOSorting} proved tight upper and
lower bounds for the I/O-complexity of sorting, computing the FFT,
permuting, and matrix transposition. Hong and Kung \cite{Pebbling}
introduced an abstract model of pebbling a computation graph to
analyze the I/O-complexity of algorithms. The vertices of the graph
that hold pebbles represent data items that are loaded into main
memory. With a limited number of pebbles available, the number of
moves needed to transfer all the pebbles from the input vertices to
the output vertices of the computation graph is the number of I/O
operations between main memory and external memory.

Interprocessor communication is a significant bottleneck in
multiprocessor architectures, and it becomes more severe as the number of
processors increases. In fact, depending on the degree of parallelism
of the problem itself, the communication time between processors
frequently limits the execution speed. Aggarwal et al.\
\cite{PRAMCommComplexity} proposed the LPRAM model for parallel
random access machines that incorporates both the computational power
and communication delay of parallel architectures. For this model,
they proved upper bounds on both computation time and communication
steps using $p$ processors for a number of algorithms, including
matrix multiplication, sorting, and computing an $n$-point FFT.

\subsection{External memory algorithms}
\label{subsecExternalMemAlg}
Vitter \cite{VitterSurvey} surveyed the state of the art in the design
and analysis of data structures and algorithms that operate on data
sets that are too large to fit in main memory. These algorithms try to
reduce the performance bottleneck of accesses to slower external
memory.

There has been considerable interest in the area of I/O-efficient
algorithms for a long time. Knuth \cite{KnuthACP3} investigated
sorting algorithms that work on files that are too large to fit in
fast internal memory. For example, when the file to be sorted is
stored on a sequential tape, a process of loading blocks of records
into internal memory where they are sorted and using the tape to
merge the sorted blocks turns out quite naturally to be more efficient
than running a sorting algorithm on the entire file.

Grossman and Silverman \cite{grossman-1973-placement} considered the
very general problem of storing records on a secondary storage device
to minimize expected retrieval time, when the probability of accessing
any record is known in advance. The authors model the pattern of
accesses by means of a parameter that characterizes the degree to
which the accesses are sequential in nature.

There has been interest in the numerical computing field in improving
the performance of algorithms that operate on large matrices
\cite{MatrixTranspose}. A successful strategy is to partition the
matrix into rectangular blocks, each block small enough to fit
entirely in main memory or cache, and operate on the blocks
independently.

The same blocking strategy has been employed for graph algorithms
\cite{awerbuch-1998-nearlinear, chiang-1995-external,
nodine-1996-blocking}. The idea is to cover an input graph with
subgraphs; each subgraph is a small diameter neighborhood of vertices
just big enough to fit in main memory. A computation on the entire
graph can be performed by loading each neighborhood subgraph into main
memory in turn, computing the final results for all vertices in the
subgraph, and storing back the results.

Gil and Itai \cite{PackTrees} studied the problem of storing a binary
tree in a virtual memory system to minimize the number of page
faults. They considered the problem of allocating the nodes of a given
binary tree (not necessarily a search tree) to virtual memory pages,
called a packing, to
optimize the cache performance for some pattern of accesses to the
tree nodes. The authors investigated the specific model for tree
accesses in which a node is accessed only via the path from the root
to that node. They presented a dynamic programming algorithm to find a
packing that minimizes the number of page faults incurred and the
number of different pages visited while accessing a node. In addition,
the authors proved that the problem of finding an optimal packing that
also uses the minimum number of pages in NP-complete, but they
presented an efficient approximation algorithm.

\subsection{Non-uniform memory architecture}
In a non-uniform memory architecture (NUMA), each processor contains a
portion of the shared memory, so access times to different parts of
the shared address space can vary, sometimes significantly.

NUMA architectures have been proposed for large-scale multiprocessor
computers. For instance, Wilson
\cite{wilson-1987-hierarchical} proposed an architecture with
hierarchies of shared buses and caches. The author proposed extensions
of cache coherency protocols to maintain cache coherency in this model
and presented simulations to demonstrate that a 128 processor computer
could be constructed using this architecture that would achieve a
substantial fraction of its peak performance.

A related architecture proposed by Hagersten et al.\
\cite{hagersten-1992-ddm}, called the Cache-Only Memory Architecture
(COMA), is similar to a NUMA in the sense that each processor holds a
portion of the shared address space. In the COMA, however, the
allocation of the shared address space among the processors can be
dynamic. All of the distributed memory is organized like large
caches. The cache belonging to each processor serves two purposes---it
caches the recently accessed data for the processor itself and also
contains a portion of the shared memory. A coherence protocol is used
to manage the caches.

\subsection{Models for non-uniform memory}
\label{subsecModels}
One motivation for a better model of computation is the desire to
model real computers more accurately. We want to to be able to design
and analyze algorithms, predict their performance, and characterize
the hardness of problems. Consequently, we want a simple, elegant
model that provides a faithful abstraction of an actual
computer. Below, we survey the theoretical models of computation that
have been proposed to model memory effects in actual computers.

The seminal paper by Aggarwal et al.\ \cite{HMM} introduced the
Hierarchical Memory Model (HMM) of computation with logarithmic memory
access cost, i.e., access to the memory location at address $a$ takes
time $\Theta(\log a)$. The HMM model seems realistic enough to model
a computer with multiple levels in the memory hierarchy. It confirms
with our intuition that successive levels in memory become slower but
bigger. Standard polynomial-time RAM algorithms can run on this HMM
model with an extra factor of at most $O(\log n)$ in the running
time. The authors showed that some algorithms can be rewritten to
reduce this factor by taking advantage of locality of reference, while
other algorithms cannot be improved asymptotically.

Aggarwal et al.\ \cite{HMBT} proposed the Hierarchical Memory model
with Block Transfer (HMBT) as a better model that incorporates the
cost of data transfer between levels in the memory hierarchy. The HMBT
model allows data to be transferred between levels in blocks in a
pipelined manner, so that it takes only constant time per unit of
memory after the initial item in the block. The authors considered
variants of the model with different memory access costs: $f(a) = \log
a$, $f(a) = a^\beta$ for $0 < \beta < 1$, and $f(a) = a$.

Aggarwal and Chandra \cite{VirtualMemAlg} proposed a model $VM_f$ for
a computer with virtual memory. The virtual memory on the $VM_f$ model
consists of a hierarchical partitioning of memory into contiguous
intervals or blocks. Some subset of the blocks at any level are stored
in faster (real) memory at any time. The blocks and sub-blocks of
virtual memory are used to model disk blocks, pages of real memory,
cache lines, etc. The authors' model for the real memory is the HMBT
model $BT_f$ in which blocks of real memory can be transferred between
memory levels in unit time per location after the initial access,
i.e., in a pipelined manner. The $VM_f$ is considered a higher-level
abstraction on which to analyze application programs, while the
running time is determined by the time taken by the underlying block
transfers. In both the models considered, the $VM_f$ and the $BT_f$,
the parameter $f$ is a memory cost function representing the cost of
accessing a location in real or virtual memory.

The Uniform Memory Hierarchy (UMH) model of computation proposed by
Alpern et al.\ \cite{UMH} incorporates a number of parameters that
model the hierarchical nature of computer memory. Like the HMBT, the
UMH model allows data transfers between successive memory levels via a
bus. The transfer cost along a bus is parameterized by the bandwidth of
the bus. Other parameters include the size of a block and the number
of blocks in each level of memory.

Regan \cite{regan-1996-linear} introduced the Block Move (BM) model of
computation that extended the ideas of the HMBT model proposed by
Aggarwal et al.\ \cite{HMBT}. The BM model allows more complex
operations such as shuffling and reversing of blocks of memory, as
well as the ability to apply other finite transductions besides
``copy'' to a block of memory. The memory-access cost of a block
transfer, similar to that in the HMBT model, is unit cost per location
after the initial access. Regan proved that different variants of the
model are equivalent up to constant factors in the memory-access
cost. He studied complexity classes for the BM model and compared them
with standard complexity classes defined for the RAM and the Turing
machine.

Two extensions of the HMBT model, the Parallel HMBT (P-HMBT) and the
pipelined P-HMBT (PP-HMBT), were investigated by Juurlink and Wijshoff
\cite{ParHMM}. In these models, data transfers between memory levels
may proceed concurrently. The authors proved tight bounds on the total
running time of several problems on the P-HMBT model with access cost
function $f(a) = \floor{\log a}$. The P-HMBT model is identical to the
HMBT model except that block transfers of data are allowed to proceed
in parallel between memory levels, and a transfer can take place only
between successive levels. In the PP-HMBT model, different block
transfers involving the same memory location can be pipelined.
The authors showed that the P-HMBT and HMBT models are incomparable in
strength, in the sense that there are problems that can be solved
faster on one model than on the other; however, the PP-HMBT model is
strictly more powerful than both the HMBT and the P-HMBT models.

A number of models have also been proposed for parallel computers with
hierarchical memory.

Valiant \cite{valiant-1989-bsp} proposed the Bulk-Synchronous Parallel
(BSP) model as an abstract model for designing and analyzing parallel
programs. The BSP model consists of \emph{components} that perform
computation and memory access tasks and a \emph{router} that delivers
messages point-to-point between the components. There is a facility to
synchronize all or a subset of components at the end of each
\emph{superstep}. The model emphasizes the separation of the task of
computation and the task of communicating between components. The
purpose of the router is to implement access by the components to
shared memory in parallel. In \cite{valiant-1990-bridging}, Valiant
argues that the BSP model can be implemented efficiently in hardware,
and therefore, it serves as both an abstract model for designing,
analyzing and implementing algorithms as well as a realistic
architecture realizable in hardware.

Culler et al.\ \cite{LogP} proposed the \textsc{LogP} model of a
distributed-memory multiprocessor machine in which processors
communicate by point-to-point messages. The performance
characteristics of the interconnection network are modeled by four
parameters $L$, $o$, $g$, and $P$: $L$ is the latency incurred in
transmitting a message over the network, $o$ is the overhead during
which the processor is busy transmitting or receiving a message, $g$
is the minimum gap (time interval) between consecutive message
transmissions or reception by a processor, and $P$ is the number of
processors or memory modules. The \textsc{LogP} model does not model
local architectural features, such as caches and pipelines, at each
processor.

For a comprehensive discussion of computational models, including
models for hierarchical memory, we refer the reader to the book by
Savage \cite{Savage}.


For the rest of this thesis, we focus on a generalization of the HMM
model due to Aggarwal et al.\ \cite{HMM} where the memory cost
function can be an arbitrary nondecreasing function, not just
logarithmic.

Now that we have a more realistic model of computation, our next goal
is to re-analyze existing algorithms and data structures, and either
prove that they are still efficient in this new model or design better
ones. Also, in the cases where we observe worse performance on the new
model, we would also like to be able to prove nontrivial lower
bounds. This leads to our primary interest in this thesis, which
studies the problem of constructing minimum-cost binary search trees
on a hierarchical memory model of computation.


\setcounter{chapter}{2}
\chapter[Algorithms for Constructing Optimum and Nearly
Optimum\newline Binary Search Trees]{Algorithms for Constructing
Optimum and Nearly Optimum Binary Search Trees}
\label{chap3main}

\section{The HMM model}
Our version of the HMM model of computation consists of a single
processor with a potentially unbounded number of memory locations with
addresses $1$, $2$, $3$, $\ldots$. We identify a memory location by
its address. A location in memory can store a finite but arbitrarily
large integer value.

The processor can execute any instruction in constant time, not
counting the time spent reading from or writing to memory. Some
instructions read operands from memory or write results into the
memory. Such instructions can address any memory location directly by
its address; this is called ``random access'' to memory, as opposed to
sequential access. At most one memory location can be accessed at a
time. The time taken to read and write a memory location is the same.

The HMM is controlled by a program consisting of a finite sequence of
instructions. The state of the HMM is defined by the sequence number
of the current instruction and the contents of memory.

In the initial state, the processor is just about to execute the first
instruction in its program. If the length of the binary representation
of the input is $n$, then memory locations $1$ through $n$ contain the
input, and all memory locations at higher addresses contain zeros.
The program is not stored in memory but encoded in the processor's
finite control.

The memory organization of the HMM model is dramatically different
from that of the RAM.  On the HMM, accessing different memory
locations may take different amounts of time.  Memory is organized in
a hierarchy, from fastest to slowest. Within each level of the
hierarchy, the cost of accessing a memory location is the same.

More precisely, the memory of the HMM is organized into a hierarchy
$M_1$, $M_2$, $\ldots$, $M_h$ with $h$ different levels, where $M_l$
denotes the set of memory locations in level $l$ for $1 \leq l \leq
h$. Let $m_l = \abs{M_l}$ be the number of memory locations in
$M_l$. The time to access every location in $M_l$ is the same.  Let
$c_l$ be the time taken to access a single memory location in
$M_l$. Without loss of generality, the levels in the memory hierarchy
are organized from fastest to slowest, so that $c_1 < c_2 < \ldots <
c_h$. We will refer to the memory locations with the lowest cost of
access, $c_1$, as the ``cheapest'' memory locations.

For an HMM, we define a memory cost function $\mu : \Nat
\rightarrow \Nat$ that gives the cost $\mu(a)$ of a single access to
the memory location at address $a$. The function $\mu$ is defined by
the following increasing step function:
\[
\mu(a) = \left\{
\begin{array}{ll}
c_1 &\text{for $0 < a \leq m_1$} \\
c_2 &\text{for $m_1 < a \leq m_1+m_2$} \\
c_3 &\text{for $m_1+m_2 < a \leq m_1+m_2+m_3$} \\
\vdots\\
c_h &\text{for $\sum_{l=1}^{h-1} m_l < a \leq \sum_{l=1}^{h} m_l$}.
\end{array}
\right.
\]
We do not make any assumptions about the relative sizes of the levels
in the hierarchy, although we expect that $m_1 < m_2 < \ldots < m_h$
in an actual computer.

A \emph{memory configuration} with $s$ locations is a sequence
$\mathcal{C}_s = \seq{n_l \mid 1 \leq l \leq h}$ where each $n_l$ is
the number of memory locations from level $l$ in the memory
hierarchy and $\sum_{l=1}^{h} n_l = s$.

The running time of a program on the HMM model consists of the time
taken by the processor to execute the instructions according to the
program and the time taken to access memory. Clearly, if even the
fastest memory on the HMM is slower than the uniform-cost memory on
the RAM, then the same program will take longer on the HMM than on the
RAM. Assume that the RAM memory is unit cost per access, and that $1
\leq c_1 < c_2 < \ldots < c_h$. Then, the running time of an
algorithm on the HMM will be at most $c_h$ times that on the RAM. An
interesting question is whether the algorithm can be redesigned to
take advantage of locality of reference so that its running time
on the HMM is less than $c_h$ times the running time on the RAM.

\section{The \HMM2 model}
The Hierarchical Memory Model with two memory levels (\HMM2) is the
special case of the general HMM model with $h=2$. In the
\HMM2, memory is organized in a hierarchy consisting of only two
levels, denoted by $\mathcal{M}_1$ and $\mathcal{M}_2$. There are
$m_1$ locations in $\mathcal{M}_1$ and $m_2$ locations in $\mathcal{M}_2$.
The total number of memory locations is $m_1 + m_2 = n$.  A single
access to any location in $\mathcal{M}_1$ takes time $c_1$, and an
access to any location in $\mathcal{M}_2$ takes time $c_2$, with $c_1
< c_2$. We will refer to the memory locations in $\mathcal{M}_1$ as
the ``cheaper'' or ``less expensive'' locations.

\section{Optimum BSTs on the HMM model}
\label{secBST4HMM}
We study the following problem for the HMM model with $n$ memory
locations and an arbitrary memory cost function $\mu : \{1$, $2$,
$\ldots$, $n\} \rightarrow \Nat$.
\begin{problem}{[\textbf{Constructing an optimum BST on the HMM}]}
\label{problem_BST4HMM}
Suppose we are given a set of $n$ keys, $x_1$, $x_2$, $\ldots$, $x_n$
in order, the probabilities $p_i$ for $1 \leq i \leq n$ that a
search argument $y$ equals $x_i$, and the probabilities $q_j$ for $0
\leq j \leq n$ that $x_{j-1} \prec y \prec x_j$.  The problem is to
construct a binary search tree $T$ over the set of keys and compute a
memory assignment function $\phi : V(T) \rightarrow \{1$, $2$,
$\ldots$, $n\}$ that assigns the (internal) nodes of $T$ to memory
locations such that the expected cost of a search is minimized.
\end{problem}

Let $\tuple{T,\phi}$ denote a potential solution to the above problem:
$T$ is the combinatorial structure of the tree, and the memory
assignment function $\phi$ maps the internal nodes of $T$ to memory
locations.

If $v$ is an internal node of $T$, then $\phi(v)$ is the address of
the memory location where $v$ is stored, and $\mu(\phi(v))$ is the
cost of a single access to $v$. If $v$ stores the key $x_i$, then we
will sometimes write $\phi(x_i)$ for $\phi(v)$. On the other hand, if
$v$ is an external node of $T$, then such a node does not actually
exist in the tree; however, it does contribute to the probability that
its parent node is accessed. Therefore, for an external node $v$, we
use $\phi(v)$ to denote the memory location where the \emph{parent} of
$v$ is stored. Let $T_v$ denote the subtree of $T$ rooted at $v$. Now
$T_v$ is a binary search tree over some subset, say $\{x_i$,
$x_{i+1}$, $\ldots$, $x_j\}$, of keys; let $w(T_v)$ denote the sum of
the corresponding probabilities: $w(T_v) = w_{i,j} =
\sum_{k=i}^{j} p_k + \sum_{k=i-1}^{j} q_k$. (If $v$ is the external
node $z_j$, we use the convention that $v$ is a subtree over the empty
set of keys from $x_{j+1}$ through $x_j$, and $w(T_v) = w_{j+1,j} =
q_j$.) Therefore, $w(T_v)$ is the probability that the search for a
key in $T$ proceeds anywhere in the subtree $T_v$.

On the HMM model, making a single comparison of the search argument
$y$ with the key $x_i$ incurs, in addition to the constant computation
time, a cost of $\mu(\phi(x_i))$ for accessing the memory location
where the corresponding node of $T$ is stored. By the cost of
$\tuple{T,\phi}$, we mean the expected cost of a search:
\begin{equation}
\label{eqn_HMMcost}
\text{cost}(\tuple{T,\phi}) = \sum_{i=1}^{n} w(T_{x_i}) \cdot
\mu(\phi(x_i)) + \sum_{j=0}^{n} w(T_{z_j}) \cdot \mu(\phi(z_j))
\end{equation}
where the first summation is over all $n$ internal nodes $x_i$ of
$T$ and the second summation is over the $n+1$ external nodes $z_j$.

Here is another way to derive the above formula---the search algorithm
accesses the node $v$ whenever the search proceeds anywhere in the
subtree rooted at $v$, and the probability of this event is precisely
$w(T_v) = w_{i,j}$. The contribution of the node $v$ to the total cost
is the probability $w(T_v)$ of accessing $v$ times the cost
$\mu(\phi(v))$ of a single access to the memory location containing
$v$.

The pair $\tuple{T^*,\phi^*}$ is an optimum solution to an instance of
problem \ref{problem_BST4HMM} if $\text{cost}(\tuple{T^*,\phi^*})$ is
minimum over all binary search trees $T$ and functions $\phi$
assigning the nodes of $T$ to memory locations. We show below in Lemma
\ref{lemma_heap} that for a given tree $T$ there is a unique function
$\phi$ that optimally assigns nodes of $T$ to memory locations.

It is easy to see that on the standard RAM model where every memory
access takes unit time, equation (\ref{eqn_HMMcost}) is equivalent to
equation (\ref{eqn_RAMcost}). Each node $v$ contributes once to
the sum on the right side of (\ref{eqn_HMMcost}) for each of its
ancestors in $T$.

\subsection{Storing a tree in memory optimally}
The following lemmas show that the problem of constructing optimum
BSTs specifically on the HMM model is interesting because
of the interplay between the two parameters---the combinatorial
structure of the tree and the memory assignment; restricted versions
of the general problem have simple solutions.

Consider the following restriction of problem \ref{problem_BST4HMM}
with the combinatorial structure of the BST $T$ fixed.
\begin{problem}
\label{prob_fixedT}
Given a binary search tree $T$ over the set of keys $x_1$ through
$x_n$, compute an optimum memory assignment function $\phi : V(T)
\rightarrow \{1$, $2$, $\ldots$, $n\}$ that assigns the nodes of $T$
to memory locations such that the expected cost of a search is
minimized.
\end{problem}

Let $\pi(v)$ denote the parent of the node $v$ in $T$; if $v$ is the
root, then let $\pi(v) = v$. Let $\phi^*$ denote an optimum memory
assignment function that assigns the nodes of $T$ to locations in
memory.

\begin{lemma}
\label{lemma_heap}
With $T$ fixed, for every node $v$ of $T$,
\[
  \mu(\phi^*(\pi(v))) \leq \mu(\phi^*(v)).
\]
In other words, for a fixed BST $T$, there exists an optimal memory
assignment function that assigns every node of $T$ to a memory
location that is no more expensive than the memory locations assigned
to its children.
\end{lemma}
\begin{proof}
Assume to the contrary that for a particular node $v$, we have
$\mu(\phi^*(\pi(v))) > \mu(\phi^*(v))$. The
contribution of $v$ and $\pi(v)$ to the total cost of the tree in the
summation (\ref{eqn_HMMcost}) is
\[
  w(T_{\pi(v)}) \mu(\phi^*(\pi(v))) + w(T_v) \mu(\phi^*(v)).
\]

The node $\pi(v)$ is accessed whenever the search proceeds anywhere in
the subtree rooted at $\pi(v)$, and likewise with $v$. Since each
$p_i, q_j \geq 0$, $\pi(v)$ is accessed at least as often as $v$,
i.e., $w(T_{\pi(v)}) \geq w(T_v)$.

Therefore, since $\mu(\phi^*(v)) < \mu(\phi^*(\pi(v)))$ by our
assumption,
\[
w(T_{\pi(v)}) \mu(\phi^*(v)) + w(T_v) \mu(\phi^*(\pi(v))) 
  \leq w(T_{\pi(v)}) \mu(\phi^*(\pi(v))) + w(T_v) \mu(\phi^*(v))
\]
so that we can swap the memory locations where $v$ and its parent
$\pi(v)$ are stored and not increase the cost of the solution.
\end{proof}

As a consequence, the root of any subtree is stored in the cheapest 
memory location among all nodes in that subtree.

\begin{lemma}
\label{lemma_greedyalg}
For fixed $T$, the optimum memory assignment function, $\phi^*$, can be
determined by a greedy algorithm. The running time of this greedy
algorithm is $O(n \log n)$ on the RAM.
\end{lemma}
\begin{proof}
It follows from Lemma \ref{lemma_heap} that under some optimum memory
assignment, the root of the tree must be assigned the cheapest
available memory location. Again from the same lemma, the next
cheapest available location can be assigned only to one of the
children of the root, and so on. The following algorithm implements
this greedy strategy.

By the \emph{weight} of a node $v$ in the tree, we mean the sum of the
probabilities of all nodes in the subtree rooted at $v$, i.e.,
$w(T_v)$. The value $w(T_v)$ can be computed for every subtree $T_v$
in linear time and stored at $v$. We maintain the set of candidates
for the next cheapest location in a heap ordered by their
weights. Among all candidates, the optimum choice is to assign the
cheapest location to the heaviest vertex. We extract this vertex,
say $u$, from the top of the heap, store it in the next available
memory location, and insert the two children of $u$ into the
heap. Initially, the heap contains just the root of the entire tree,
and the algorithm continues until the heap is empty.

This algorithm performs $n$ insertions and $n$ deletions on a heap
containing at most $n$ elements. Therefore, its running time on the
uniform-cost RAM model is $O(n \log n)$.
\end{proof}

\subsection[Constructing an optimum tree when the memory assignment
is \newline fixed]{Constructing an optimum tree when the memory
assignment is fixed}
Consider the following restriction of problem
\ref{problem_BST4HMM} where the memory assignment function $\phi$ is
given.
\begin{problem}
\label{problem_fixedmem}
Suppose each of the keys $x_i$, for $1 \leq i \leq n$, is assigned
\emph{a priori} a fixed location $\phi(x_i)$ in memory. Compute the
structure of a binary search tree of minimum cost where every node
$v_i$ of the tree corresponding to key $x_i$ is stored in memory
location $\phi(x_i)$.
\end{problem}

\begin{lemma}
\label{lemma_optimality}
Given a fixed assignment of keys to memory locations, i.e., a function
$\phi$ from the set of keys (equivalently, the set of nodes of any BST
$T$) to the set of memory locations, the BST $T^*$ of minimum cost can
be constructed by a dynamic programming algorithm. The running time of
this algorithm is $O(n^3)$ on the RAM.
\end{lemma}
\begin{proof}
The principle of optimality clearly applies here so that a BST is
optimum if and only if each subtree is optimum. The standard dynamic
programming algorithm proceeds as follows:

Let $\text{cost}(T^*_{i,j})$ denote the cost of an optimum BST over the keys
$x_i$, $x_{i+1}$, $\ldots$, $x_j$ and the corresponding
probabilities $p_i$, $p_{i+1}$, $\ldots$, $p_j$ and $q_{i-1}$,
$q_i$, $\ldots$, $q_j$, given the fixed memory assignment $\phi$. By
the principle of optimality,
\begin{align}
\label{eqn_optimality}
\text{cost}(T^*_{i,j}) &= w_{i,j} \cdot \mu(\phi(x_k)) 
+ \min_{i \leq k \leq j}\Paren{\text{cost}(T^*_{i,k-1}) +
\text{cost}(T^*_{k+1,j})}
 &\text{for $i \leq j$} \nonumber\\
\text{cost}(T^*_{i+1,i}) &= w_{i+1,i} = q_i.
\end{align}
Recall that $w_{i,j}$ is the probability that the root of this
subtree is accessed, and $\mu(\phi(x_k))$ is the cost of a single
access to the memory location $\phi(x_k)$ where $x_k$ is stored.

Notice that this expression is equivalent to equation
(\ref{eqn_recurrenceBST}) except for the multiplicative factor
$\mu(\phi(x_k))$. Therefore, \textsc{algorithm K1} from section
\ref{n3algorithm} can be used to construct the optimum binary search
tree efficiently, given an assignment of keys to memory locations.
\end{proof}

In general, it does not seem possible to use a monotonicity
principle to reduce the running time to $O(n^2)$, as in
\textsc{algorithm K2} of section \ref{n2algorithm}.

\subsection{Naive algorithm}
A naive algorithm for problem \ref{problem_BST4HMM} is to try every
possible mapping of keys to memory locations. Lemma
\ref{lemma_optimality} guarantees that we can then use dynamic
programming to construct an optimum binary search tree for that memory
assignment. We select the minimum-cost tree over all possible memory
assignment functions.

There are 
\[
    \binom{n}{m_1, m_2, \ldots, m_h}
\]
such mappings from $n$ keys to $n$ memory locations with $m_1$ of the
first type, $m_2$ of the second type, and so on. The multinomial
coefficient is maximized when $m_1 = m_2 = \cdots = m_{h-1} =
\Floor{n/h}$. The dynamic programming algorithm takes $O(n^3)$ time to
compute the optimum BST for each fixed memory assignment. Hence, the
running time of the naive algorithm is
\begin{align}
O\Paren{\frac{n!}{\Paren{\frac{n}{h}!}^h} \cdot n^3} 
&= O\Paren{\frac{\sqrt{2 \pi n} (n/e)^n}{(\sqrt{2 \pi
(n/h)}((n/h)/e)^{(n/h)})^h} \cdot n^3} \nonumber\\
&\qquad\text{using Stirling's approximation} \nonumber\\
&= O\Paren{\frac{\sqrt{2 \pi n}}{(\sqrt{2 \pi (n/h)})^h} \cdot h^n \cdot n^3} \nonumber\\
&= O\Paren{\frac{h^{(h/2)}}{(2 \pi n)^{(h-1)/2}} \cdot h^n \cdot n^3}
\nonumber\\
&= O\Paren{\frac{h^{n+h/2} \cdot n^{3-(h-1)/2}}{(2 \pi)^{(h-1)/2}}} \nonumber\\
&= O(h^n \cdot n^3).
\end{align}

Unfortunately, the above algorithm is inefficient and therefore
infeasible even for small values of $n$ because its running time is
exponential in $n$. We develop much more efficient algorithms in the
following sections.

\subsection{A dynamic programming algorithm: \textsc{algorithm Parts}}
\label{secn2hp2algorithm}
A better algorithm uses dynamic programming to construct optimum
subtrees bottom-up, like \textsc{algorithm K1} from section
\ref{n3algorithm}. Our new algorithm, \textsc{algorithm Parts}, constructs
an optimum subtree $T^*_{i,j}$ for each $i$, $j$, such that $1 \leq i
\leq j \leq n$ and for every memory configuration $\seq{n_1, n_2,
\ldots, n_h}$ consisting of the $j-i+1$ memory locations available at
this stage in the computation. For each possible choice $x_k$ for the
root of the subtree $T_{i,j}$, there are at most $j-i+2 \leq n+1$
different ways to partition the number of available locations in each
of $h-1$ levels of the memory hierarchy between the left and right
subtrees of $x_k$. (Since the number of memory locations assigned to
any subtree equals the number of nodes in the subtree, we have the
freedom to choose only the number of locations from any $h-1$ levels
because the number of locations from the remaining level is then
determined.)

We modify \textsc{algorithm K1} from section \ref{n3algorithm} as
follows. \textsc{algorithm K1} builds larger and larger optimum
subtrees $T^*_{i,j}$ for all $i$, $j$ such that $1 \leq i \leq j
\leq n$. For every choice of $i$ and $j$, the algorithm iterates through
the $j-i+1$ choices for the root of the subtree from among $\{x_i$,
$x_{i+1}$, $\ldots$, $x_j\}$. The left subtree of $T^*_{i,j}$ with
$x_k$ at the root is a BST, say $T^{(L)}$, over the keys $x_i$ through
$x_{k-1}$, and the right subtree is a BST, say $T^{(R)}$, over the
keys $x_{k+1}$ through $x_j$.

The subtree $T_{i,j}$ has $j-i+1$ nodes. Suppose the number of memory
locations available for the subtree $T_{i,j}$ from each of the memory
levels is $n_l$ for $1 \leq l \leq h$, where $\sum_{l=1}^{h} n_l =
j-i+1$. There are
\begin{align*}
\binom{(j-i+1)+h-1}{h-1} &= \binom{j-i+h}{h-1} \\
 &= O\Paren{\frac{(n+h)^{h-1}}{(h-1)!}} \\
 &= O\Paren{\frac{2^{h-1}}{(h-1)!} n^{h-1}}
     &\text{since $h \leq n$}
\end{align*}
different ways to partition $j-i+1$ objects into $h$ parts without
restriction, and therefore, at most as many different memory
configurations with $j-i+1$ memory locations. (There are likely to be
far fewer different memory configurations because there are at most
$m_1$ memory locations from the first level, at most $m_2$ from the
second, and so on, in any configuration.)

Let $\lambda$ be the smallest integer such that $n_{\lambda} > 0$;
in other words, the cheapest available memory location is from memory
level $\lambda$.

For every choice of $i$, $j$, and $k$, there are at most $\min\{k-i+1,
n_{\lambda}\} \leq n$ different choices for the number of memory
locations from level $\lambda$ to be assigned to the left
subtree, $T^{(L)}$. This is because the left subtree with $k-i$ nodes
can be assigned any number from zero to $\max\{k-i,n_{\lambda}-1\}$
locations from the first available memory level,
$\mathcal{M}_{\lambda}$. (Only at most $n_{\lambda}-1$ locations from
$\mathcal{M}_{\lambda}$ are available after the root $x_k$ is stored
in the cheapest available location.) The remaining locations from
$\mathcal{M}_{\lambda}$ available to the entire subtree are assigned
to the right subtree, $T^{(R)}$. Likewise, there are at most
$\min\{k-i+1,n_{\lambda+1}+1\}
\leq n+1$ different choices for the number of ways to partition the
available memory locations from the next memory level
$\mathcal{M}_{\lambda+1}$ between the left and right subtrees, and so
on. In general, the number of memory locations from the memory level
$l$ assigned to the left subtree, $n^{(L)}_l$, ranges from $0$ to at
most $n_l$. Correspondingly, the number of memory locations from the
level $l$ assigned to the right subtree $n^{(R)}_l$ is $n_l -
n^{(L)}_l$.

We modify \textsc{algorithm K1} by inserting $h - \lambda \leq h-1$
more nested loops that iterate through every such way to partition the
available memory locations from $\mathcal{M}_{\lambda}$ through
$\mathcal{M}_{h-1}$ between the left and right subtrees of $T_{i,j}$
for a fixed choice of $x_k$ as the root.

\begin{figure}
\begin{center}\fbox{
\begin{algorithm}
\textul{\textsc{algorithm Parts}:}
\\ \comment{Initialization}
\\ for $i$ := $0$ to $n$ \+
\\   Let $\mathcal{C}_0$ be the empty memory configuration $\seq{0,0,\ldots,0}$
\\   $C[i+1,i,\mathcal{C}_0] \leftarrow q_i$;
\\   $R[i+1,i,\mathcal{C}_0] \leftarrow \NIL$; \-
\\
\\ for $d$ := $0$ to $n-1$ \+
\\   \comment{Construct optimum subtrees with $d+1$ nodes.}
\\   for each memory configuration $\mathcal{C}$ of size $d+1$ \+
\\     for $i$ := $1$ to $n-d$ \+
\\       $j \leftarrow i+d$
\\       $C[i,j,\mathcal{C}] \leftarrow \infty$
\\       $R[i,j,\mathcal{C}] \leftarrow \NIL$
\\       for $k$ := $i$ to $j$ \+
\\         \comment{Number of nodes in the left and right subtrees.}
\\         $l \leftarrow k-i$ \qquad \comment{number of nodes in the
left subtree}
\\         $r \leftarrow j-k$ \qquad \comment{number of nodes in the
right subtree}
\\
\\         Call \textsc{procedure Partition-Memory} (figure \ref{fig_procPartnMem}) to compute
\\         the optimum way to partition the available memory locations.
\end{algorithm}
}\end{center}
\caption{\textsc{algorithm Parts}}
\label{fig_algorithmParts}
\hrule
\end{figure}
\begin{figure}
\begin{center}\fbox{
\begin{algorithm}
\textul{\textsc{procedure Partition-Memory}:}
\\ Let $\mathcal{C} \equiv \seq{n_1, n_2, \ldots, n_h}$.
\\ Let $\lambda$ be the smallest integer such that $n_{\lambda} > 0$.
\\
\\ for $n^{(L)}_{\lambda}$ := $0$ to $n_{\lambda}$ \+
\\   for $n^{(L)}_{\lambda+1}$ := $0$ to $n_{\lambda+1}$ \+
\\     $\ddots$
\\     for $n^{(L)}_{h-1}$ := $0$ to $n_{h-1}$ \+
\\        $n^{(L)}_h \leftarrow l - \sum_{i=1}^{h-1} n^{(L)}_i$
\\        $n^{(R)}_{\lambda} \leftarrow n_{\lambda} - n^{(L)}_{\lambda}$
\\        $n^{(R)}_{\lambda+1} \leftarrow n_{\lambda+1} - n^{(L)}_{\lambda+1}$
\\        $\vdots$
\\        $n^{(R)}_{h-1} \leftarrow n_{h-1} - n^{(L)}_{h-1}$
\\        $n^{(R)}_h \leftarrow r - \sum_{i=1}^{h-1} n^{(R)}_i$
\\
\\                 Use one cheap location for the root, i.e.,
\\                   $n^{(L)}_{\lambda} \leftarrow n^{(L)}_{\lambda}-1$
\\                   $n^{(R)}_{\lambda} \leftarrow
n^{(R)}_{\lambda}-1$
\\
\\                 Let $\mathcal{C}^L = \seq{0, \ldots, 0, n^{(L)}_{\lambda}, n^{(L)}_{\lambda+1},
\ldots, n^{(L)}_h}$.
\\                 Let $\mathcal{C}^R =
\seq{0,\ldots,0,(n_{\lambda}-1) - n^{(L)}_{\lambda},
n_{\lambda+1} - n^{(L)}_{\lambda+1},\ldots,n_h - n^{(L)}_h}$.
\\
\\                 Let $T'$ be the tree with $x_k$ at the root, and the
left and right children
\\                 are given by $R[i,k-1,\mathcal{C}^L]$ and $R[k+1,j,\mathcal{C}^R]$ respectively.
\\                 i.e., $T'$ is the tree
\\                 \pstree[treesep=0.5in, levelsep=0.5in]
{\Tcircle{$x_k$}}{\Trect{$T[i,k-1,\mathcal{C}^L]$}
\Trect{$T[k+1,j,\mathcal{C}^R]$}}
\\
\\                 \comment{Let $c'$ be the cost of $T'$.}
\\                 \comment{The root of $T'$ is stored in a location
of cost $c_{\lambda}$.}
\\                 $C' \leftarrow c_{\lambda} \cdot w_{i,j} +
C[i,k-1,\mathcal{C}^L] + C[k+1,j,\mathcal{C}^R]$
\\
\\                 if $C' < C[i,j,\mathcal{C}]$ \+
\\                   $R[i,j,\mathcal{C}] \leftarrow \tuple{k, \mathcal{C}^L}$
\\                   $C[i,j,\mathcal{C}] \leftarrow C'$ \-\-\-
\end{algorithm}
}\end{center}
\caption{\textsc{procedure Partition-Memory}}
\label{fig_procPartnMem}
\hrule
\end{figure}

Just like \textsc{algorithm K1}, \textsc{algorithm Parts} of figure
\ref{fig_algorithmParts} constructs arrays $R$ and $C$, each indexed
by the pair $i$, $j$, such that $1 \leq i \leq j \leq n$, and the
memory configuration $\mathcal{C}$ specifying the numbers of memory
locations from each of the $h$ levels available to the subtree
$T_{i,j}$. Let $\mathcal{C} = \seq{n_1, n_2, \ldots, n_h}$. The array
entry $R[i,j,\mathcal{C}]$ stores the pair $\tuple{k,\mathcal{C}^L}$,
where $k$ is the index of the root of the optimum subtree $T^*_{i,j}$
for memory configuration $\mathcal{C}$, and $\mathcal{C}^L$ is the
optimum memory configuration for the left subtree. In other words,
$\mathcal{C}^L$ specifies for each $l$ the number of memory locations
$n^{(L)}_l$ out of the total $n_l$ locations from level $l$ available
to the subtree $T_{i,j}$ that are assigned to the left subtree.  The
memory configuration $\mathcal{C}^R$ of the right subtree is
automatically determined: the number of memory locations $n^{(R)}_l$
from level $l$ that are assigned to the right subtree is $n_l -
n^{(L)}_l$, except that one location from the cheapest memory level
available is consumed by the root.

The structure of the optimum BST and the optimum memory assignment
function is stored implicitly in the array $R$. Let
$T[i,j,\mathcal{C}]$ denote the implicit representation of the optimum
BST over the subset of keys from $x_i$ through $x_j$ for memory
configuration $\mathcal{C}$. If $R[1,n,\mathcal{C}]
= \seq{k, \mathcal{C}'}$, then the root of the entire tree is $x_k$
and it is stored in the cheapest available memory location of cost
$c_{\lambda}$. The left subtree is over the subset of keys $x_1$
through $x_{k-1}$, and the memory configuration for the left subtree
is $\mathcal{C}' = \seq{0,\ldots,0,n_{\lambda}', n_{\lambda+1}',
\ldots, n_h'}$. The right subtree is over the subset of keys $x_{k+1}$
through $x_n$, and the memory configuration for the right subtree is
$\seq{0,\ldots,0,(n_{\lambda}-1) - n_{\lambda}', n_{\lambda+1} -
n_{\lambda+1}', \ldots, n_h - n_h'}$.

In \textsc{algorithm Parts}, there are $3+(h-1) = h+2$ nested loops each
of which iterates at most $n$ times, in addition to the loop that
iterates over all possible memory configurations of size $d+1$ for $0
\leq d \leq n-1$. Hence, the running time of the
algorithm is
\begin{align}
O\Paren{\frac{2^{h-1}}{(h-1)!} n^{h-1} \cdot n^{h+2}}
 &= O\Paren{\frac{2^{h-1}}{(h-1)!} \cdot n^{2h+1}}.
\end{align}

\subsection{Another dynamic programming algorithm: \textsc{algorithm Trunks}}
\label{secnmhalgorithm}
In this subsection, we develop another algorithm that iteratively
constructs optimum subtrees $T^*_{i,j}$ over larger and larger subsets
of keys. Fix an $i$ and $j$ with $1 \leq i \leq j \leq n$ and $j-i=d$,
and a memory configuration $\mathcal{C}_{s+1} = \seq{n_1, n_2, \ldots,
n_{h-1}, n_h}$ consisting of $s+1$ memory locations from the first
$h-1$ levels of the memory hierarchy and none from the last level,
i.e., $n_1 + n_2 + \cdots + n_{h-1} = s+1$ and $n_h = 0$. At iteration
$s+1$, we require an optimum subtree, over the subset of keys from
$x_i$ through $x_j$, with $s$ of its nodes assigned to
memory locations from the first $h-1$ levels of the memory hierarchy
and the remaining $(j-i+1)-s$ nodes stored in the most expensive
locations. Call the subtree induced by the nodes stored in the first
$h-1$ memory levels the \emph{trunk} (short for ``truncated'') of the
tree. (Lemma \ref{lemma_heap} guarantees that the trunk will also be a
tree, and the root of the entire tree is also the root of the
trunk. So, in fact, a trunk with $s+1$ nodes of a tree is obtained by
pruning the tree down to $s+1$ nodes by recursively deleting leaves.)
We require the optimum subtree $T^*_{1,n}$ with $\sum_{r=1}^{h-1} m_r
= n - m_h$ nodes in the trunk, all of which are assigned to the $n-m_h$
locations in the cheapest $h-1$ memory levels. Recall that $m_l$ is
the number of memory locations in memory level $l$ for $1 \leq l \leq h$.

\textsc{algorithm Trunks} in figure \ref{fig_algorithmTrunks}
constructs a table indexed by $i$, $j$, and $\mathcal{C}_{s+1}$. There
are $\binom{n}{2}$ different choices of $i$ and $j$ such that $1 \leq
i \leq j \leq n$. Also, there are
\[
  \binom{(s+1)+(h-1)-1}{h-2} = \binom{s+h-1}{h-2}
\]
different ways to partition $s+1$ objects into $h-1$ parts without
restriction, and therefore, at most as many different memory
configurations with $s+1$ memory locations from the first $h-1$ memory
levels. (As mentioned earlier, there are likely to be far fewer
different memory configurations because there are restrictions on the
number of memory locations from each level in any configuration.)

For every value of $k$ from $i$ to $j$ and every $t$ from $0$ to $s$,
we construct a subtree with $x_k$ at the root and $t$ nodes in the
trunk of the left subtree (the \emph{left trunk}) and $s-t$ nodes in
the trunk of the right subtree (the \emph{right trunk}).

\begin{figure}
\begin{center}\fbox{
\begin{algorithm}
\textul{\textsc{algorithm Trunks}:}
\\ Initially, the optimum subtree $T^*_{i,j}$ is unknown for all $i$,
$j$,
\\ except when the subtree fits entirely in memory level $M_h$,
\\ in which case the optimum subtree is the one
\\ computed by \textsc{algorithm K2} during the initialization phase.
\\
\\ for $d$ := $0$ to $n-1$ \+
\\   for $i$ := $1$ to $n-d$ \+
\\     $j \leftarrow i+d$
\\     \comment{Construct an optimum BST over the subset of keys from
$x_i$ through $x_j$.}
\\     for $k$ := $i$ to $j$ \+
\\       \comment{Choose $x_k$ to be the root of this subtree.}
\\
\\       for $s$ := $1$ to $n-m_h-1$ \+
\\         \comment{Construct a BST with $s$ nodes in its trunk.}
\\
\\         For every memory configuration $\mathcal{C}_s$ of size $s$ \+
\\
\\           for $t$ := $0$ to $s$ \+
\\             \comment{The left trunk has $t$ nodes.}
\\             For every choice of $t$ out of the $s$ memory locations
\\             in $\mathcal{C}_s$ to assign to the left subtree. \+
\\               Let $T'$ be the BST over the subset of keys from
$x_i$ through $x_j$
\\               with $x_k$ at the root,
\\               $t$ nodes in the trunk of the left subtree, and
\\               $s-t$ nodes in the trunk of the right subtree.
\\
\\               The left subtree of $T'$ is the previously computed
\\               optimum subtree over the keys $x_i$ through $x_{k-1}$
\\               with $t$ nodes in its trunk, and the right subtree of
$T'$
\\               is the previously computed optimum subtree over the
\\               keys $x_{k+1}$ through $x_j$ with $s-t$ nodes in its trunk.
\\
\\               If the cost of $T'$ is less than that of the
minimum-cost
\\               subtree found so far, then record $T'$ as the new
\\               optimum subtree.
\end{algorithm}
}\end{center}
\caption{\textsc{algorithm Trunks}}
\label{fig_algorithmTrunks}
\hrule
\end{figure}

By Lemma \ref{lemma_heap}, the root of the subtree $x_k$ is always
stored in the cheapest available memory location. There are at most
$\binom{s}{t}$ ways to select $t$ out of the remaining $s$ memory
locations to assign to the left trunk. (In fact, since the $s$ memory
locations are not necessarily all distinct, there are likely to be far
fewer ways to do this.) As $t$ iterates from $0$ through $s$, the
total number of ways to partition the available $s$ memory locations
and assign them to the left and right trunks is at most
\[
	\sum_{t=0}^{s} \binom{s}{t} = 2^s.
\]

When all the nodes of the subtree are stored in memory
locations in level $h$ (the base case when $s=0$), an optimum subtree
$T^*_{i,j}$ is one constructed by \textsc{algorithm K2} from section
\ref{n2algorithm}. Therefore, in an initial phase, we execute
\textsc{algorithm K2} to construct, in $O(n^2)$ time, all optimum
subtrees $T^*_{i,j}$ that fit entirely within one memory level, in
particular, the last and most expensive memory level.

The total running time of the dynamic programming algorithm is,
therefore,
\[
  O\Paren{n^2 + \sum_{d=0}^{n-1} \sum_{i=1}^{n-d} \sum_{k=i}^{i+d}
    \sum_{s=0}^{n-m_h-1} \binom{s+h-1}{h-2} \cdot 2^s }.
\]
Let
\[
  f(n) = \sum_{s=0}^{n-m_h-1} \binom{s+h-1}{h-2} \cdot 2^s.
\]
By definition,
\[
  f(n) \leq \sum_{s=0}^{n-m_h-1} \frac{(s+h-1)^{h-2}}{(h-2)!} 
              \cdot 2^s
       = \frac{1}{(h-2)!} \sum_{s=0}^{n-m_h-1} (s+h-1)^{h-2}
              \cdot 2^s.
\]
Thus, $f(n)$ is bounded above by the sum of a geometric series whose
ratio is at most $2 \cdot (n-m_h-1+h-1)$. Hence, we have
\begin{align*}
f(n) &\leq \frac{1}{(h-2)!} \cdot \frac{2^{n-m_h} (n-m_h+h-2)^{n-m_h} 
           - 1}{2(n-m_h+h-2) - 1} \\
     &= O\Paren{\frac{2^{n-m_h} \cdot (n-m_h+h)^{n-m_h}}{(h-2)!}}.
\end{align*}
Therefore, the running time of the algorithm is
\begin{align}
&O\Paren{\sum_{d=0}^{n-1} \sum_{i=1}^{n-d} \sum_{k=i}^{j=i+d}
\frac{2^{n-m_h} \cdot (n-m_h+h)^{n-m_h}}{(h-2)!}} \nonumber\\
&\qquad = O\Paren{\frac{2^{n-m_h} \cdot (n-m_h+h)^{n-m_h}}{(h-2)!}
\sum_{d=0}^{n-1} \sum_{i=1}^{n-d} (d+1) } \nonumber\\
&\qquad = O\Paren{\frac{2^{n-m_h} \cdot (n-m_h+h)^{n-m_h} \cdot n^3}{(h-2)!}}
\label{time_algorithmTrunks}.
\end{align}

\textsc{algorithm Trunks} is efficient when $n-m_h$ and $h$ are both small. 
For instance, consider a memory organization in which the memory cost
function grows as the tower function defined by:
\begin{align*}
\text{tower}(0) &= 1 \\
\text{tower}(i+1) &= 2^{\text{tower}(i)} =
\left.{2^{2^{.^{.^{.^{2}}}}}}\right\} \text{($i+1$ times)}
&\text{for all $i \geq 1$}.
\end{align*}
If $\mu(a) = \text{tower}(a)$ is the memory cost function, then
$\sum_{r=1}^{h-1} m_r = n - m_h < \lg\Paren{\sum_{r=1}^{h} m_r} = 
\lg n$, and $h = \log^* n$. For all practical purposes, $\log^* n$ is
a small constant; therefore, the running time bound of equation
\ref{time_algorithmTrunks} is almost a polynomial in $n$.

\subsection{A top-down algorithm: \textsc{algorithm Split}}
\label{sec2nalgorithm}
Suppose there are $n$ distinct memory costs, or $n$ levels in the
memory hierarchy with one location in each level. A top-down recursive
algorithm to construct an optimum BST has to decide at each step in
the recursion how to partition the available memory locations between
the left and right subtrees. Note that the number of memory locations
assigned to the left subtree determines the number of keys in the left
subtree, and therefore identifies the root. So, for example, if $k$ of
the available $n$ memory locations are assigned to the left subtree,
then there are $k$ keys in the left subtree, and hence, the root of
the tree is $x_{k+1}$.

At the top level, the root is assigned the cheapest memory location.
Each of the remaining $n-1$ memory locations can be assigned to either
the left or the right subtree, so that $k$ of the $n-1$ locations are
assigned to the left subtree and $n-1-k$ locations to the right
subtree for every $k$ such that $0 \leq k \leq n-1$. Thus, there are
$2^{n-1}$ different ways to partition the available $n-1$ memory
locations between the two subtrees of the root. The algorithm proceeds
recursively to compute the left and right subtrees.

The asymptotic running time of the above algorithm is given by the
recurrence
\[
  T(n) = 2^{n-1} + \max_{0 \leq k \leq n-1} \Set{T(k) + T(n-1-k)}.
\]
Now, $T(n)$ is at least $2^{n-1}$, which is a convex function, and
$T(n)$ is a monotonically increasing function of $n$. Therefore, a
simple inductive argument shows that $T(n)$ itself is convex, so that
it achieves the maximum value at either $k=0$ or $k=n-1$. At $k=0$,
$T(n) = 2^{n-1} + T(0) + T(n-1)$ which is the same value as at
$k=n-1$. Therefore,
\begin{align}
T(n) &\leq 2^{n-1} + T(0) + T(n-1) \nonumber\\
     &= \sum_{i=0}^{n-1} 2^i \nonumber\\
     &= 2^n - 1 \nonumber\\
     &= O(2^n).
\end{align}

\section{Optimum BSTs on the \HMM2 model}
\label{secBST4HMM2}
In this section, we consider the problem of constructing and storing
an optimum BST on the \HMM2 model. Recall that the \HMM2 model
consists of $m_1$ locations in memory level $\mathcal{M}_1$, each of
cost $c_1$, and $m_2$ locations in memory level $\mathcal{M}_2$, each
of cost $c_2$, with $c_1 < c_2$.

\subsection{A dynamic programming algorithm}
\label{secdynprog4hmm2}
In this section, we develop a hybrid dynamic programming algorithm to
construct an optimum BST. Recall that \textsc{algorithm K2} of section
\ref{secOptBST4RAM} constructs an optimum BST for the uniform-cost
RAM model in $O(n^2)$ time. It is an easy observation that the
structure of an optimum subtree that fits entirely in one memory level
is the same as that of the optimum subtree on the uniform-cost RAM
model. Therefore, in an initial phase of our hybrid algorithm,
we construct optimum subtrees with at most $\max{m_1,m_2}$ nodes that
fit in the largest memory level. In phase II, we construct larger
subtrees.

Recall from equation (\ref{eqn_recurrenceBST}) that on the uniform-cost
RAM model the cost $c(i,j)$ of an optimum BST over the subset of keys
from $x_i$ through $x_j$ is given by the recurrence
\begin{align*}
c(i+1,i) &= w_{i+1,i} = q_i \\
c(i,j) &= w_{i,j} + \min_{i \leq k \leq j}\ \Paren{c(i,k-1) + c(k+1,j)} 
            &\text{when $i \leq j$}
\end{align*}

On the \HMM2 model, the cost of an optimum BST $T^*_{i,j}$ over the
same subset of keys is
\begin{align}
c(i+1,i,n_1,n_2) &= q_i \nonumber\\
c(i,j,n_1,n_2) &= \mu(\phi(x_k)) \cdot w_{i,j} \nonumber\\
	       &\qquad {}+ \min_{i \leq k \leq j \atop
                                 0 \leq n^{(L)}_1 < n_1}
  \Paren{c(i,k-1,n^{(L)}_1,n^{(L)}_2) + c(k+1,j,n^{(R)}_1,n^{(R)}_2)}
\end{align}
where
\begin{itemize}
\item the root $x_k$ is stored in memory location $\phi(x_k)$ of cost
$\mu(\phi(x_k))$;
\item out of the $n_1$ cheap locations available to the subtree,
$n^{(L)}_1$ are given to the left subtree and $n^{(R)}_1$ are given to
the right subtree;
\item the $n_2$ expensive locations available are assigned as
$n^{(L)}_2$ to the left subtree and $n^{(R)}_2$ to the right subtree;
\item if $n_1 > 0$, then $x_k$ is stored in a location of cost $c_1$,
and $n^{(L)}_1 + n^{(R)}_1 = n_1 - 1$ and $n^{(L)}_2 + n^{(R)}_2 =
n_2$;
\item otherwise, $n_1 = 0$ and $n_2 = j-i+1$, so $x_k$ is stored in a
location of cost $c_2$, and the entire subtree is stored in the second
memory level; the optimum subtree $T^*_{i,j}$ is the same as
the optimum one on the RAM model constructed during phase I.
\end{itemize}

The first phase of the algorithm, \textsc{procedure TL-phase-I} constructs
arrays $C$ and $R$, where $C[i,j]$ is the cost of an optimum BST (on
the uniform-cost model) over the subset of keys from $x_i$ through
$x_j$; $R[i,j]$ is the index of the root of such an optimum BST.

The second phase, \textsc{procedure TL-phase-II}, constructs arrays $c$ and
$r$, such that $c[i,j,n_1,n_2]$ is the cost of an optimum BST over the
subset of keys from $x_i$ through $x_j$ with $n_1$ and $n_2$ available
memory locations of cost $c_1$ and $c_2$ respectively, and $n_1 + n_2 =
j-i+1$; $r[i,j,n_1,n_2]$ is the index of the root of such an optimum
BST.

The structure of the tree can be retrieved in $O(n)$ time from the
array $r$ at the end of the execution of \textsc{algorithm TwoLevel}.

\subsubsection{\textsc{algorithm TwoLevel}}
\textsc{algorithm TwoLevel} first calls \textsc{procedure TL-phase-I}. Recall
that \textsc{procedure TL-phase-I} constructs all subtrees
$T_{i,j}$ that contain few enough nodes to fit entirely in any one
level in the memory hierarchy, specifically the largest level.
Entries in table $R[i,j]$ are filled by \textsc{procedure TL-phase-I}.

\textsc{procedure TL-phase-II} computes optimum subtrees where $n_1$ and
$n_2$ are greater than zero. Therefore, prior to invoking algorithm
\textsc{TL-phase-II}, \textsc{algorithm TwoLevel} initializes the entries
in table $r[i,j,n_1,n_2]$ when $n_1 = 0$ and when $n_2 = 0$ from the
entries in table $R[i,j]$.

\begin{figure}
\begin{center}\fbox{
\begin{algorithm}
\textul{\textsc{algorithm TwoLevel}:}
\\  Call \textsc{procedure TL-phase-I} (figure \ref{fig_procTLphase-I})
\\  If either $m_1=0$ or $m_2=0$, then we are done.
\\  Otherwise, \+
\\    Initialize, for all $i$, $j$ such that $1 \leq i \leq j \leq n$: \+
\\      $r[i,j,0,j-i+1] \leftarrow R[i,j]$
\\      $r[i,j,j-i+1,0] \leftarrow R[i,j]$
\\      $c[i,j,0,j-i+1] \leftarrow c_2 \cdot C[i,j]$
\\      $c[i,j,j-i+1,0] \leftarrow c_1 \cdot C[i,j]$ \-
\\  Call \textsc{procedure TL-phase-II} (figure \ref{fig_procTLphase-II})
\end{algorithm}
}\end{center}
\caption{\textsc{algorithm TwoLevel}}
\label{fig_algorithmTwoLevel}
\hrule
\end{figure}

\subsubsection{Procedure \textsc{TL-phase-I}}
\textsc{procedure TL-phase-I} is identical to \textsc{algorithm K2} from
section \ref{n2algorithm} except that the outermost loop involving $d$
iterates only $\max\{m_1$, $m_2\}$ times in \textsc{procedure
TL-phase-I}. \textsc{procedure TL-phase-I} computes optimum subtrees in a
bottom-up fashion. It fills entries in the tables $C[i,j]$ and
$R[i,j]$ by diagonals, i.e., in the order of increasing $d = j-i$. The
size of the largest subtree that fits entirely in one memory level is
$\max\{m_1$, $m_2\}$, corresponding to $d = \max\{m_1$, $m_2\} - 1$.

For every $i,j$ with $j-i = d$, \textsc{TL-phase-I} computes the cost of a
subtree $T'$ with $x_k$ at the root for all $k$, such that $R[i,j-1]
\leq k \leq R[i+1,j]$. Note that $(j-1)-i = j-(i+1) = d-1$; therefore,
entries $R[i,j-1]$ and $R[i+1,j]$ are already available during this
iteration of the outermost loop. The optimum choice for the root of
this subtree is the value of $k$ for which the cost of the subtree is
minimized.

\begin{figure}
\begin{center}\fbox{
\begin{algorithm}
\textul{\textsc{procedure TL-phase-I}:}
\\  \comment{Initialization phase.}
\\  for $i$ := $0$ to $n$ \+
\\    $C[i+1,i] \leftarrow w_{i+1,i} = q_i$
\\    $R[i+1,i] \leftarrow \NIL$ \-
\\
\\  for $d$ := $0$ to $\max\set{m_1,m_2} - 1$ \+
\\    for $i$ := $1$ to $n-d$ \+
\\      $j \leftarrow i + d$
\\      \comment{Number of nodes in this subtree: $j-i+1 = d+1$.}
\\      $C[i,j] \leftarrow \infty$
\\      $R[i,j] \leftarrow \NIL$
\\      for $k$ := $R[i,j-1]$ to $R[i+1,j]$ \+
\\        ($\star$) \fbox{
          $T'$ is the tree
          \pstree[treesep=0.5in, levelsep=0.5in]
{\Tcircle{$x_k$}}{\Trect{$T[i,k-1]$} \Trect{$T[k+1,j]$}}
          }
\\        $C' \leftarrow w_{i,j} + C[i,k-1] + C[k+1,j]$
\\        if $C' < C[i,j]$ \+
\\          $R[i,j] \leftarrow k$
\\          $C[i,j] \leftarrow C'$ \-\-\-\-
\end{algorithm}
}\end{center}
\caption{\textsc{procedure TL-phase-I}}
\label{fig_procTLphase-I}
\hrule
\end{figure}

\subsubsection{Procedure \textsc{TL-phase-II}}
\textsc{procedure TL-phase-II} is an implementation of \textsc{algorithm Parts}
in section \ref{secn2hp2algorithm} for the special case when
$h=2$. \textsc{procedure TL-phase-II} also constructs increasingly larger
optimum subtrees in an iterative fashion.  The additional complexity
in this algorithm arises from the fact that for each possible choice
of root $x_k$ of the subtree $T_{i,j}$, there are also a number of
different ways to partition the available cheap locations between the
left and right subtrees of $x_k$.

There are $m_1$ cheap locations and $m_2$ expensive locations
available to store the subtree $T_{i,j}$. If $m_1 \geq 1$, then the
root $x_k$ is stored in a cheap location. The remaining cheap
locations are partitioned into two, with $n^{(L)}_1$ locations
assigned to the left subtree and $n^{(R)}_1$ locations assigned to the
right subtree. $n^{(L)}_2$ and $n^{(R)}_2$ denote the number of
expensive locations available to the left and right subtrees
respectively. Since the algorithm constructs optimum subtrees in
increasing order of $j-i$, the two table entries
$r[i,k-1,n^{(L)}_1,n^{(L)}_2]$ and $r[k+1,j,n^{(R)}_1,n^{(R)}_2]$ are
already available during the iteration when $j-i=d$ because $(k-1)-i
< d$ and $j-(k+1) < d$.

\begin{figure}
\begin{center}\fbox{
\begin{algorithm}
\textul{\textsc{procedure TL-phase-II}:}
\\ for $d$ := $\min\set{m_1,m_2}$ to $n-1$ \+
\\   for $n_1$ := $0$ to $\min\set{m_1,d+1}$ \+
\\     $n_2 \leftarrow (d+1) - n_1$
\\     for $i$ := $1$ to $n-d$ \+
\\       $j \leftarrow i+d$
\\       $c[i,j,n_1,n_2] \leftarrow \infty$
\\       $r[i,j,n_1,n_2] \leftarrow \NIL$
\\       for $k$ := $i$ to $j$ \+
\\         \comment{Number of nodes in the left and right subtrees.}
\\         $l \leftarrow k-1$
\\         $r \leftarrow n-k$
\\         if $n_1 \geq 1$ \+
\\           Use one cheap location for the root;
\\           \comment{Now, there are only $n_1-1$ cheap locations available.}
\\           for $n^{(L)}_1$ := $\max\set{0,(n_1-1)-r}$ to $\min\set{l,(n_1-1)}$ \+
\\             $n^{(L)}_2 \leftarrow l - n^{(L)}_1$
\\             $n^{(R)}_1 \leftarrow (n_1-1) - n^{(L)}_1$
\\             $n^{(R)}_2 \leftarrow r - n^{(R)}_1$
\\             $(\star)$ \fbox{
               $T' \leftarrow
\pstree[treesep=0.5in, levelsep=0.5in]
{\Tcircle{x_k}}{\Trect{T[i,k-1,n^{(L)}_1,n^{(L)}_2]}
\Trect{T[k+1,j,n^{(R)}_1,n^{(R)}_2]}}$
               }
\\
\\             $c' \leftarrow c_1 \cdot w_{i,j} +
c[i,k-1,n^{(L)}_1,n^{(L)}_2] + c[k+1,j,n^{(R)}_1,n^{(R)}_2]$
\\             if $c' < c[i,j,n_1,n_2]$ \+
\\               $r[i,j,n_1,n_2] \leftarrow k$
\\               $c[i,j,n_1,n_2] \leftarrow c'$ \-\-\-\-
\end{algorithm}
}\end{center}
\caption{\textsc{procedure TL-phase-II}}
\label{fig_procTLphase-II}
\hrule
\end{figure}

\subsubsection{Correctness of \textsc{algorithm TwoLevel}}
\textsc{algorithm TwoLevel} calls \textsc{procedure TL-phase-I} and
\textsc{procedure TL-phase-II}, which implement dynamic programming to
build larger and larger subtrees of minimum cost. The principle of
optimality clearly applies to the problem of constructing an optimum
tree---every subtree of an optimal tree must also be optimal given the
same number of memory locations of each kind. Therefore,
\textsc{algorithm TwoLevel} correctly computes an optimum BST over the
entire set of keys.

\subsubsection{Running time of \textsc{algorithm TwoLevel}}
The running time of \textsc{algorithm TwoLevel} is proportional to the
number of times overall that the lines marked with a star ($\star$) in
\textsc{TL-phase-I} and \textsc{TL-phase-II} are executed.

Let $m = \min\{m_1$, $m_2\}$ be the size of the smaller of the two
memory levels. The number of times that the line in algorithm
\textsc{TL-phase-I} marked with a star ($\star$) is executed is
\begin{align*}
\sum_{d=0}^{n-m} \sum_{i=1}^{n-d} \Paren{R[i+1,j] - R[i,j-1] + 1} 
&= \sum_{d=0}^{n-m} \Paren{R[n-d+1,n+1] - R[1,d-1] + n - d} \nonumber\\
&\leq \sum_{d=0}^{n-m} 2n \nonumber\\
&= 2n (n-m+1) \nonumber\\
&= O(n (n-m)).
\end{align*}

The number of times that the line ($\star$) in \textsc{procedure TL-phase-II}
is executed is at most
\[
  \sum_{d=m}^{n-1} \sum_{n_1=0}^{\min\{m_1,d+1\}} \sum_{i=1}^{n-d} \sum_{k=i}^{i+d} m.
\]
A simple calculation shows that the two summations involving $d$ and
$i$ iterate $O(n-m)$ times each, the summation over $n_1$ iterates
$O(n)$ times, and the innermost summation has $O(n)$ terms, so
that the number of times that the starred line is executed is $O(m n^2
(n-m)^2)$.

Therefore, the total running time of \textsc{algorithm TwoLevel} is 
\begin{equation}
T(n,m) = O(n (n-m) + m n^2 (n-m)^2) = O(m n^2 (n-m)^2).
\end{equation}
In general, $T(n,m) = O(n^5)$, but $T(n,m) = o(n^5)$ if $m = o(n)$, and
$T(n,m) = O(n^4)$ if $m = O(1)$, i.e., the smaller level in memory
has only a constant number of memory locations. This case would arise
in architectures in which the faster memory, such as the primary
cache, is limited in size due to practical considerations such as
monetary cost and the cost of cache coherence protocols.

\subsection{Constructing a nearly optimum BST}
\label{secApproxBST}
In this section, we consider the problem of constructing a BST on the
\HMM2 model that is close to optimum.

\subsubsection{An approximation algorithm}
The following top-down recursive algorithm, \textsc{algorithm
Approx-BST} of figures \ref{fig_algorithmApproxBST} and
\ref{fig_algorithmApproxBST2}, is due to Mehlhorn
\cite{Mehlhorn}. Its analysis is adapted from the same source.
The intuition behind \textsc{algorithm Approx-BST} is to choose the
root $x_k$ of the subtree $T_{i,j}$ so that the weights $w_{i,k-1}$
and $w_{k+1,j}$ of the left and right subtrees are as close to equal
as possible. In other words, we choose the key $x_k$ to be the root
such that $\Abs{w_{i,k-1} - w_{k+1,j}}$ is as small as possible. Then,
we recursively construct the left and right subtrees.

Once the tree $\tilde{T}$ has been constructed by the above heuristic,
we optimally assign the nodes of $\tilde{T}$ to memory
locations using Lemma \ref{lemma_greedyalg} in $O(n \log n)$
additional time.

Algorithm \textsc{Approx-BST} implements the above heuristic.  The
parameter $l$ represents the depth of the recursion; initially $l=0$,
and $l$ is incremented by one whenever the algorithm recursively calls
itself. The parameters $\text{low}_l$ and $\text{high}_l$ represent
the lower and upper bounds on the range of the probability
distribution spanned by the keys $x_i$ through $x_j$. Initially,
$\text{low}_l = 0$ and $\text{high}_l = 1$ because the keys $x_1$
through $x_n$ span the entire range $[0,1]$. Whenever the root $x_k$
is chosen, according to the above heuristic, to lie in the middle of
this range, i.e., $mid_l = (\text{low}_l + \text{high}_l)/2$, the span
of the keys in the left subtree is bounded by $[\text{low}_l,
\text{med}_l]$ and the span of the keys in the right subtree is
bounded by $[med_l, \text{high}_l]$. These are the ranges passed as
parameters to the two recursive calls of the algorithm.

Define
\begin{align}
\label{def_si1}
	s_0	&= \frac{q_0}{2} \nonumber\\
	s_i	&= s_{i-1} + \frac{q_{i-1}}{2} + p_i + \frac{q_{i}}{2}
			&\text{for $1 \leq i \leq n$}
\end{align}
By definition,
\begin{align}
\label{def_si2}
	s_i	&= \frac{q_0}{2} + \sum_{k=1}^{i} p_k
			+ \sum_{k=1}^{i-1} q_k + \frac{q_i}{2} \nonumber\\
		&= w_{1,i} - \frac{q_0}{2} - \frac{q_i}{2}
\end{align}
Therefore,
\begin{align}
s_j - s_{i-1} &= w_{1,j} - w_{1,i-1} + \frac{q_{i-1}}{2} -
\frac{q_j}{2} \nonumber\\
          &= w_{i,j} + \frac{q_{i-1}}{2} - \frac{q_j}{2} 
             &\text{by definition \ref{eqn_w_ij}}
\end{align}
In Lemma \ref{lemma_params} below, we show that at each level in
the recursion, the input parameters to \textsc{Approx-BST}$()$ satisfy
$\text{low}_l \leq s_{i-1} \leq s_j \leq \text{high}_l$.

\begin{figure}
\begin{center}\fbox{
\begin{algorithm}
\textul{\textsc{Approx-BST}$(i, j, l, \text{low}_l, \text{high}_l)$:}
\\	$med_l \leftarrow (\text{low}_l + \text{high}_l) / 2$;
\\	Case 1: (the base case)
\\	if $i = j$ \+
\\		Return the tree with three nodes consisting of $x_i$
at the root 
\\              and the external nodes $z_{i-1}$ and $z_j$ as the left and right subtrees respectively:
\\ 		\pstree{\Tcircle{$x_i$}}{\pstree{\Trect{$z_{i-1}$}}{}
\pstree{\Trect{$z_i$}}{}} \-
\\	Otherwise, if $i \neq j$, then find $k$ satisfying all the following three conditions:\+
\\		(i) $i \leq k \leq j$
\\		(ii) either $k = i$, or $k>i$ and $s_{k-1} \leq med_l$
\\		(iii) either $k = j$, or $k<j$ and $s_k \geq med_l$\-
\\              (Lemma \ref{lemma_kexists} guarantees that such a $k$
always exists.)
\\
\\      \hfill\textsl{(Continued in figure \ref{fig_algorithmApproxBST2})}
\end{algorithm}
}\end{center}
\caption{\textsc{algorithm Approx-BST}}
\label{fig_algorithmApproxBST}
\hrule
\end{figure}
\begin{figure}
\begin{center}\fbox{
\begin{algorithm}
        \hfill\textsl{(Continued from figure \ref{fig_algorithmApproxBST})}
\\
\\	\textbf{Case 2a:}
\\	if $k = i$ \+
\\		Return the tree with $x_i$ at the root, the external
node $z_{i-1}$ as the left subtree, \+
\\                and the recursively constructed subtree $T_{i+1,j}$
as the right subtree:
\\		\pstree{\Tcircle{$x_i$}}{\pstree{\Trect{$z_{i-1}$}}{}
\pstree{\Trect{\textsc{Approx-BST}$(i+1, j, l+1, med_l, \text{high}_l)$}}{}}
\-\-
\\
\\	\textbf{Case 2b:}
\\	if $k = j$ \+
\\		Return the tree with $x_j$ at the root, the external
node $z_j$ as the right subtree, \+
\\                and the recursively constructed subtree $T_{i,j-1}$
as the left subtree:
\\		\pstree{\Tcircle{$x_j$}}{\pstree{\Trect{\textsc{Approx-BST}$(i,
j-1, l+1, \text{low}_l, med_l)$}}{}
\pstree{\Trect{$z_j$}}{}} 
\-\-
\\
\\	\textbf{Case 2c:}
\\	if $i < k < j$ \+
\\		Return the tree with $x_k$ at the root,
\\              and recursively construct the left and right subtrees,
\\              $T_{i,k-1}$ and $T_{k+1,j}$ respectively: \+
\\                call \textsc{Approx-BST}$(i, k-1, l+1,
\text{low}_l, med_l)$ recursively \+
\\                  to construct the left subtree. \-
\\                call \textsc{Approx-BST}$(k+1, j, l+1,
med_l, \text{high}_l)$ recursively \+
\\                  to construct the right subtree. \-
\end{algorithm}
}\end{center}
\caption{\textsc{algorithm Approx-BST} (cont'd.)}
\label{fig_algorithmApproxBST2}
\hrule
\end{figure}

\subsubsection{Analysis of the running time}
We prove that the running time of algorithm \textsc{Approx-BST} is
$O(n)$. Clearly, the space complexity is also linear.

The running time $t(n)$ of \textsc{algorithm Approx-BST} can be expressed by
the recurrence
\begin{align}
t(n) &= s(n) + \max_{1 \leq k \leq n} \Bracket{t(k-1) + t(n-k)}
\end{align}
where $s(n)$ is the time to compute the index $k$ satisfying
conditions (i), (ii), and (iii) given in the algorithm, and $t(k-1)$
and $t(n-k)$ are the times for the two recursive calls.

We can implement the search for $k$ as a binary search. Initially,
choose $r = \floor{(i+j)/2}$. If $s_r \geq med_l$, then $k \leq r$,
otherwise $k \geq r$, and we proceed recursively. Since this binary
search takes $O(\log (j-i)) = O(\log n)$ time, the overall running
time of algorithm \textsc{Approx-BST} is
\begin{align*}
t(n) &= O(\log n) + \max_{1 \leq k \leq n} \Bracket{t(k-1) + t(n-k)} \\
     &\leq O(\log n) + t(0) + t(n-1) \\
     &= O(n \log n).
\end{align*}

However, if we use exponential search and then binary search to
determine the value of $k$, then the overall running time can be
reduced to $O(n)$ as follows. Intuitively, an exponential search
followed by a binary search finds the correct value of $k$ in $O(\log
(k-i))$ time instead of $O(\log (j-i))$ time.

Initially, choose $r = \floor{(i+j)/2}$. Now, if $s_r \geq med_l$ we
know $k \leq r$, otherwise $k > r$.

Consider the case when $k \in \{i$, $i+1$, $i+2$, $\ldots$,
$r=\floor{(i+j)/2}\}$. An exponential search for $k$ in this interval
proceeds by trying all values of $k$ from $i$, $i+2^0$, $i+2^1$,
$i+2^2$, and so on up to $i+2^{\ceil{\lg (r-i)}} \geq r$. Let $g$ be the
smallest integer such that $s_{i+2^g} \geq med_l$, i.e., $i+2^{g-1} <
k \leq i+2^g$, or $2^g \geq k-i > 2^{g-1}$. Hence, $\lg (k-i) > g-1$,
so that the number of comparisons made by this exponential search is
$g < 1 + \lg (k-i)$. Now, we determine the exact value of $k$ by a
binary search on the interval $i+2^{g-1}+1$ through $i+2^g$, which
takes $\lg (2^g - 2^{g-1})) + 1 < g + 1 < \lg (k-i) + 2$ comparisons.

Likewise, when $k \in \{r+1$, $r+2$, $\ldots$, $j\}$, a search for
$k$ in this interval using exponential and then binary search takes
$\lg (j-k) + 2$ comparisons.

Therefore, the time $s(n)$ taken to determine the value of $k$ is at
most $d(2 + \lg(\min\set{k-i,j-k}))$, where $d$ is a constant.

Hence, the running time of algorithm \textsc{Approx-BST} is
proportional to
\[
t(n) = \max_{1 \leq k \leq n} \Paren{t(k-1) + t(n-k) + d(2 + \lg
\min\set{k,n-k}) + f}
\]
where $f$ is a constant. By the symmetry of the expression $t(k-1) +
t(n-k)$, we have
\begin{equation}
\label{eqn_rec2}
t(n) \leq \max_{1 \leq k \leq (n+1)/2} \Paren{t(k-1) + t(n-k) + d(2
+ \lg k) + f}.
\end{equation}

We prove that $t(n) \leq (3d+f)n - d \lg(n+1)$ by induction on
$n$. This is clearly true for $n=0$. Applying the induction hypothesis
in the recurrence in equation (\ref{eqn_rec2}), we have
\begin{align*}
t(n) &\leq \max_{1 \leq k \leq (n+1)/2} (3d+f)(k-1) - d \lg k +
(3d+f)(n-k) \\
     &\qquad {}- d \lg (n-k+1) + d(2 + \lg k) + f) \\
  &= (3d+f)(n-1) + \max_{1 \leq k \leq (n+1)/2} \Paren{-d \lg (n-k+1) +
2d+f} \\
  &= (3d+f) n + \max_{1 \leq k \leq (n+1)/2} \Paren{-d \lg (n-k+1) - d}.
\end{align*}
The expression $-d(1 + \lg(n-k+1))$ is always negative and its value
is maximum in the range $1 \leq k \leq (n+1)/2$ at $k=(n+1)/2$. Therefore,
\begin{align*}
t(n) &\leq (3d+f) n - d(1 + \lg ((n+1)/2)) \\
     &= (3d+f) n - d \lg (n+1).
\end{align*}

Hence, the running time of algorithm \textsc{Approx-BST} is
$O(t(n)) = O(n)$.

Of course, if we choose to construct an optimal memory assignment for
$\tilde{T}$, then the total running time is $O(n + n \log n) = O(n
\log n)$.

\subsubsection{Quality of approximation}
Let $\tilde{T}$ denote the binary search tree constructed by
algorithm \textsc{Approx-BST}. In the rest of this section, we prove
an upper bound on how much the cost of $\tilde{T}$ is worse than the
cost of an optimum BST. The following analysis applies whether we
choose to construct an optimal memory assignment or to use the
heuristic of algorithm \textsc{Approx-BST}.

We now derive an upper bound on the cost of the tree, $\tilde{T}$,
constructed by algorithm \textsc{Approx-BST}.

Let $\delta(x_i)$ denote the depth of the internal node $x_i$, $1 \leq i
\leq n$, and let $\delta(z_j)$ denote the depth of the external node $z_j$,
$0 \leq j \leq n$ in $\tilde{T}$.  (Recall that the depth of a node
is the number of nodes on the path from the root to that node; the
depth of the root is $1$.)

\begin{lemma}
\label{lemma_kexists}
If the parameters $i$, $j$, $\text{low}_l$, and $\text{high}_l$ to
\textsc{Approx-BST}$()$ satisfy
\[
  \text{low}_l \leq s_{i-1} \leq s_j \leq \text{high}_l,
\]
then a $k$ satisfying conditions (i), (ii), and (iii) stated in the
algorithm always exists.
\end{lemma}
\begin{proof}
If $s_i \geq \text{med}_l$, then choosing $k = i$ satisfies 
conditions (i), (ii), and (iii). Likewise, if $s_{j-1} \leq
\text{med}_l$, then $k=j$ satisfies all the conditions. Otherwise, if
$s_i < \text{med}_l < s_{j-1}$, then since $s_i \leq s_{i+1} \leq
\cdots \leq s_{j-1} \leq s_j$, consider the first $k$, with $k > i$,
such that $s_{k-1} \leq \text{med}_l$ and $s_k \geq
\text{med}_l$. Then $k < j$ and $s_k \geq \text{med}_l$, and this
value of $k$ satisfies all three conditions.
\end{proof}

\begin{lemma}
\label{lemma_lowhigh}
The parameters of a call to \textsc{Approx-BST} satisfy
\[
  \text{high}_l = \text{low}_l + 2^{-l}.
\]
\end{lemma}
\begin{proof}
The proof is by induction on $l$. The initial call to
\textsc{Approx-BST} with $l=0$ has $\text{low}_l = 0$ and
$\text{high}_l = 1$. Whenever the algorithm recursively constructs the
left subtree $T_{i,k-1}$ in cases 2b and 2c, we have $\text{low}_{l+1}
= \text{low}_l$ and $\text{high}_{l+1} = \text{med}_l = (\text{low}_l
+ \text{high}_l)/2 = (2 \text{low}_l + 2^{-l})/2 = \text{low}_l +
2^{-l-1} = \text{low}_{l+1} + 2^{-(l+1)}$. On the other hand, whenever
the algorithm recursively constructs the right subtree $T_{k+1,j}$, in
cases 2a and 2c, we have $\text{high}_{l+1} = \text{high}_l$ and
$\text{low}_{l+1} = \text{med}_l = \text{high}_{l+1} - 2^{-(l+1)}$.
\end{proof}

\begin{lemma}
\label{lemma_params}
The parameters of a call
\textsc{Approx-BST}$(i,j,l,\text{low}_l,\text{high}_l)$
satisfy
\[
  \text{low}_l \leq s_{i-1} \leq s_j \leq \text{high}_l.
\]
\end{lemma}
\begin{proof}
The initial call is \textsc{Approx-BST}$(1, n, 1, 0, 1)$. Therefore,
$s_{i-1} = s_0 = q_0 \geq 0$ and $s_j = s_n = 1 - q_0/2 - q_n/2 \leq
1$. Thus, the parameters to the initial call to
\textsc{Approx-BST}$()$ satisfy the given condition.

The rest of the proof follows by induction on $l$. In case 2a, the
algorithm chooses $k=i$ because $s_i \geq \text{med}_l$, and
recursively constructs the right subtree over the subset of keys from
$x_{i+1}$ through $x_j$. Therefore, we have $\text{low}_{l+1} =
\text{med}_l \leq s_i \leq s_j \leq \text{high}_l =
\text{high}_{l+1}$.

In case 2b, the algorithm chooses $k=j$ because $s_{j-1} \leq
\text{med}_l$, and then recursively constructs the left subtree over
the subset of keys from $x_i$ through $x_{j-1}$. Therefore, we have
$\text{low}_{l+1} = \text{low}_l \leq s_{i-1} \leq s_{j-1} \leq
\text{med}_l = \text{high}_{l+1}$.

In case 2c, \textsc{algorithm Approx-BST} chooses $k$ such that
$s_{k-1} \leq \text{med}_l \leq s_k$ and $i < k < j$. Therefore,
during the recursive call to construct the left subtree over the
subset of keys from $x_i$ through $x_{k-1}$, we have
$\text{low}_{l+1} = \text{low}_l \leq s_{i-1} \leq s_{k-1} \leq
\text{med}_l = \text{high}_{l+1}$. During the recursive call to
construct the right subtree over the subset of keys from $x_{k+1}$
through $x_j$, we have
$\text{low}_{l+1} = \text{med}_l \leq s_k \leq s_j \leq \text{high}_l
= \text{high}_{l+1}$.
\end{proof}

\begin{lemma}
\label{lemma_depthisl}
During a call to \textsc{Approx-BST} with parameter $l$, 
if an internal node $x_k$ is created, then $\delta(x_k) = l+1$,
and if an external node $z_k$ is created, then $\delta(z_k) = l+2$.
\end{lemma}
\begin{proof}
The proof is by a simple induction on $l$. The root, at depth $1$, is
created when $l=0$. The recursive calls to construct the left and
right subtrees are made with the parameter $l$ incremented by $1$.
The depth of the external node created in cases 2a and 2b is one more
than the depth of its parent, and therefore equal to $l+2$.
\end{proof}

\begin{lemma}
\label{lemma_boundondepth1}
For every internal node $x_k$ such that $1 \leq k \leq n$,
\[
  p_k \leq 2^{-\delta(x_k)+1}
\]
and for every external node $z_k$ such that $0 \leq k \leq n$,
\[
  q_k \leq 2^{-\delta(z_k)+2}.
\]
\end{lemma}
\begin{proof}
Let the internal node $x_k$ be created during a call to
\textsc{Approx-BST}$(i,j,\text{low}_l,\text{high}_l)$. Then,
\begin{align*}
s_j - s_{i-1} &\leq \text{high}_l - \text{low}_l &\text{by Lemma
                \ref{lemma_params}} \\
              &= 2^{-l} &\text{by Lemma \ref{lemma_lowhigh}} \\
s_j - s_{i-1} &= w_{1,j} - \frac{q_j}{2} - w_{1,i-1} + \frac{q_{i-1}}{2}
              &\text{by definition of $s_{i-1}$ and $s_j$} \\
              &\geq p_k &\text{because $i \leq k \leq j$.}
\end{align*}
Therefore, by Lemmas \ref{lemma_params} and 
\ref{lemma_lowhigh}, for the internal node $x_k$ ($i \leq k \leq j$)
with probability $p_k$, we have $p_k \leq s_j - s_{i-1} \leq 2^{-l} =
2^{-\delta(x_k)+1}$ by Lemma \ref{lemma_depthisl}.

Likewise, for the external node $z_k$ ($i-1 \leq k \leq j$) with
corresponding probability of access $q_k$, we have
\begin{align*}
s_j - s_{i-1} &= \sum_{r=i}^{j} p_r + \sum_{r=i-1}^{j-1} q_r +
\frac{q_j}{2} - \frac{q_{i-1}}{2}
 &\text{by definition \ref{def_si2}} \\
&= \sum_{r=i}^{j} p_r + \frac{q_{i-1}}{2} + \sum_{r=i}^{j-1} q_r +
\frac{q_j}{2}
\end{align*}
Therefore, since $i-1 \leq k \leq j$, we have
\begin{align*}
q_k &\leq 2(s_j - s_{i-1}) \\
    &\leq 2(\text{high}_l - \text{low}_l) &\text{by Lemma \ref{lemma_params}} \\
    &= 2^{-l+1} &\text{by Lemma \ref{lemma_lowhigh}} \\
    &= 2^{-\delta(z_k)+2} &\text{by Lemma \ref{lemma_depthisl}}.
\end{align*}
\end{proof}

\begin{lemma}
\label{lemma_boundondepth2}
For every internal node $x_k$ such that $1 \leq k \leq n$,
\[
  \delta(x_k) \leq \Floor{\lg \Paren{\frac{1}{p_k}}} + 1
\]
and for every external node $z_k$ such that $0 \leq k \leq n$,
\[
  \delta(z_k) \leq \Floor{\lg \Paren{\frac{1}{q_k}}} + 2.
\]
\end{lemma}
\begin{proof}
Lemma \ref{lemma_boundondepth1} shows that $p_k \leq
2^{-\delta(x_k)+1}$. Taking logarithms of both sides to the base $2$,
we have $\lg p_k \leq - \delta(x_k) + 1$; therefore, $\delta(x_k) \leq
- \lg p_k + 1 = \lg (1/p_k) + 1$. Since the depth of $x_k$ is an
integer, we conclude that $\delta(x_k) \leq \floor{\lg (1/p_k)} +
1$. Likewise, for external node $z_k$, $\delta(z_k) \leq \floor{\lg
(1/q_k)} + 2$.
\end{proof}

Now we derive an upper bound on $\text{cost}(\tilde{T})$.  Let $H$
denote the entropy of the probability distribution $q_0$, $p_1$,
$q_1$, $\ldots$, $p_n$, $q_n$ \cite{InfoTheory}, i.e.,
\begin{equation}
\label{def_entropy}
  H = \sum_{i=1}^{n} p_i \lg \frac{1}{p_i} + \sum_{j=0}^{n} q_j \lg
\frac{1}{q_j}.
\end{equation}

If all the internal nodes of $\tilde{T}$ were stored in the expensive
locations, then the cost of $\tilde{T}$ would be at most
\begin{align}
\label{eqn_pesscountT}
&\sum_{i=1}^{n} c_2 p_i \delta(x_i) + \sum_{j=0}^{n} c_2 q_j
 (\delta(z_j)-1) \nonumber\\
&\qquad\leq c_2 \Paren{\sum_{i=1}^{n} p_i \Paren{\lg \frac{1}{p_i} + 1} +
 \sum_{j=0}^{n} q_j \Paren{\lg \frac{1}{q_j} + 1}} \nonumber\\
&\qquad\qquad\text{by Lemma \ref{lemma_boundondepth2}} \nonumber\\
&\qquad= c_2 \Paren{\Paren{\sum_{i=1}^{n} p_i \lg \frac{1}{p_i} +
 \sum_{j=0}^{n} q_j \lg \frac{1}{q_j}} + \Paren{\sum_{i=1}^{n} p_i +
 \sum_{j=0}^{n} q_j}} \nonumber\\
&\qquad= c_2 (H + 1) \nonumber\\
&\qquad\qquad\text{by definition \ref{def_entropy}
and because $\sum_{i=1}^{n} p_i + \sum_{j=0}^{n} q_j = 1$.}
\end{align}

\subsubsection{Lower bounds}
The following lower bounds are known for the cost of an optimum binary
search tree $T^*$ on the standard uniform-cost RAM model.

\begin{theorem}[Mehlhorn \cite{MehlhornLB}]
\label{thm_lbMehlhorn}
\[
  \text{cost}(T^*) \geq \frac{H}{\lg 3}
\]
\end{theorem}

\begin{theorem}[De~Prisco, De~Santis \cite{DePrisco}]
\label{thm_lbDePrisco1}
\[
  \text{cost}(T^*) \geq H - 1 - \Paren{\sum_{i=1}^{n} p_i} (\lg \lg
  (n+1) - 1).
\]
\end{theorem}

\begin{theorem}[De~Prisco, De~Santis \cite{DePrisco}]
\label{thm_lbDePrisco2}
\label{thm_bestlb}
\[
  \text{cost}(T^*) \geq H + H \lg H - (H+1) \lg (H+1).
\]
\end{theorem}

The lower bounds of Theorems \ref{thm_lbMehlhorn} and
\ref{thm_lbDePrisco2} are expressed only in terms of $H$, the entropy
of the probability distribution. The smaller the entropy, the tighter
the bound of Theorem \ref{thm_lbMehlhorn}. Theorem
\ref{thm_lbDePrisco2} improves on Mehlhorn's lower bound for $H
\gtrapprox 15$. Theorem \ref{thm_lbDePrisco1} assumes knowledge of
$n$, and proves a lower bound better than that of Theorem
\ref{thm_lbMehlhorn} for large enough values of $H$.

\subsubsection{Approximation bound}
\begin{corollary}
\label{cor_approxbound}
The algorithm \textsc{Approx-BST} constructs the tree $\tilde{T}$ such that
\[
\text{cost}(\tilde{T}) - \text{cost}(T^*)
\leq (c_2 - c_1) H + c_1 ((H+1) \lg (H+1) - H \lg H) + c_2.
\]
\end{corollary}
\begin{proof}
Theorem \ref{thm_bestlb} immediately implies a lower bound of $c_1 (H
+ H \lg H - (H+1) \lg (H+1))$ on the cost of $T^*$. The result then
follows from equation (\ref{eqn_pesscountT}).
\end{proof}

For large enough values of $H$, $H+1 \approx H$ so that $\lg (H+1)
\approx \lg H$; hence, $(H+1) \lg (H+1) - H \lg H \approx \lg
H$. Thus, we have
\begin{equation}
\label{eqn_approx4largeH}
\text{cost}(\tilde{T}) - \text{cost}(T^*) \lessapprox 
(c_2 - c_1) H + c_1 (\lg H).
\end{equation}
When $c_1 = c_2 = 1$ as in the uniform-cost RAM model, equation
(\ref{eqn_approx4largeH}) is the same as the approximation bound
obtained by Mehlhorn \cite{Mehlhorn}.

\setcounter{chapter}{3}
\chapter{Conclusions and Open Problems}
\label{chap4openconclusions}

\section{Conclusions}
\label{secConclusions}

\begin{figure}[t]
\begin{center}
\begin{tabular}{|l|l|l|l|}\hline
\textbf{Model} & \textbf{Algorithm} & \textbf{Section} &
\textbf{Running time} \\\hline
HMM &
\textsc{algorithm Parts} \ignore{(Dynamic programming)} &
\ref{secn2hp2algorithm} &
$O\Paren{\frac{2^{h-1}}{(h-1)!} \cdot n^{2h+1}}$ \\\hline
HMM &
\textsc{algorithm Trunks} \ignore{(Hybrid)} &
\ref{secnmhalgorithm} &
$O(2^{n-m_h} \cdot (n-m_h+h)^{n-m_h} \cdot n^3 / (h-2)!)$ \\\hline
HMM &
\textsc{algorithm Split} \ignore{(Top-down)} &
\ref{sec2nalgorithm} &
$O(2^n)$ \\\hline
\HMM2 &
\textsc{algorithm TwoLevel} \ignore{(Dynamic programming)} &
\ref{secdynprog4hmm2} &
$O(m n^2 (n-m)^2)$ \\\hline
\end{tabular}
\end{center}
\caption{Summary of results}
\label{table_summary1}
\hrule
\end{figure}

The table of figure \ref{table_summary1} summarizes our results for
the problem of constructing an optimum binary search tree over a set
of $n$ keys and the corresponding probabilities of access, on the
general HMM model with an arbitrary number of levels in the memory
hierarchy and on the two-level \HMM2 model. Recall that $h$ is the
number of memory levels, and $m_l$ is the number of memory locations
in level $l$ for $1 \leq l \leq h$.

We see from table \ref{table_summary1} that \textsc{algorithm Parts} is
efficient when $h$ is a small constant. The running time of
\textsc{algorithm Parts} is independent of the sizes of the different
memory levels. On the other hand, the running time of
\textsc{algorithm Trunks} is polynomial in $n$ precisely when $n - m_h =
\sum_{l=1}^{h-1} m_l$ is a constant, even if $h$ is large. Therefore,
for instance, \textsc{algorithm Parts} would be appropriate for a
three-level memory hierarchy, where the binary search tree has to be
stored in cache, main memory, and on disk. \textsc{algorithm Trunks} would
be more efficient when the memory hierarchy consists of many levels
and the last memory level is extremely large. This is
because \textsc{algorithm Trunks} uses the speed-up technique due to Knuth
\cite{KnuthBST, KnuthACP3} and Yao \cite{Yao} to take advantage of the
fact that large subtrees of the BST will in fact be stored entirely in
the last memory level.

When $h$ is large and $n-m_h$ is not a constant, the relatively simple
top-down algorithm, \textsc{algorithm Split}, is the most efficient. In
particular, when $h = \Omega(n/\log n)$, it is faster than
\textsc{algorithm Parts}.

For the \HMM2 model, we have the hybrid algorithm, $\textsc{algorithm
TwoLevel}$, with running time $O(n(n-m) + m n^2 (n-m)^2)$, where $m =
\min\{m_1$, $m_2\}$ is the size of the smaller of the two memory
levels ($m \leq n/2$). Procedure \textsc{TL-phase-II} of \textsc{algorithm
TwoLevel} is an implementation of \textsc{algorithm Parts} for a special
case. The running time of \textsc{algorithm TwoLevel} is $O(n^5)$ in the
worst case, the same as the worst-case running time of \textsc{algorithm
Parts} for $h=2$. However, if $m = o(n)$, then
\textsc{algorithm TwoLevel} outperforms \textsc{algorithm Parts}; in
particular, if $m = \Theta(1)$, then the running time of
\textsc{algorithm TwoLevel} is $O(n^4)$.

None of our algorithms depend on the actual costs of accessing a
memory location in different levels. We state as an open problem below
whether it is possible to take advantage of knowledge of the relative
costs of memory accesses to design a more efficient algorithm for
constructing optimum BSTs.

For the problem of approximating an optimum BST on the \HMM2 model, we
have a linear-time algorithm, \textsc{algorithm Approx-BST} of section
\ref{secApproxBST}, that constructs the tree $\tilde{T}$ such that
\begin{align*}
\text{cost}(\tilde{T}) - \text{cost}(T^*)
\leq (c_2 - c_1) H + c_1 ((H+1) \lg (H+1) - H \lg H) + c_2
\end{align*}
where $\text{cost}(T^*)$ is the cost of an optimum BST.

\section{Open problems}
\label{secOpenProblems}

\subsection{Efficient heuristics}
We noted above that our algorithms do not assume any relationship
between the costs $c_l$ of accessing a memory location in level $l$,
$1 \leq l \leq h$. It should be possible to design an algorithm, more
efficient than any of the algorithms in this thesis, that takes
advantage of knowledge of the memory costs to construct an optimum
binary search tree. The memory cost function $\mu(a) =
\Theta(\log a)$ would be especially interesting in this context.

\subsection{NP-hardness}
\begin{conjecture}
The problem of constructing a BST of minimum cost on the HMM
with $h=\Omega(n)$ levels in the memory hierarchy is NP-hard.
\end{conjecture}

The dynamic programming algorithm, \textsc{algorithm Parts}, of section
\ref{secn2hp2algorithm} runs in time $O(n^{h+2})$, which is efficient
only if $h=\Theta(1)$. We conjecture that when $h=\Omega(n)$, the
extra complexity of the number of different ways to store the keys in
memory, in addition to computing the structure of an optimum BST,
makes the problem hard.

\subsection{An algorithm efficient on the HMM}
Although we are interested in the problem of constructing a BST and
storing it in memory such that the cost on the HMM is minimized, we
analyze the running times of our algorithms on the RAM model. It would
be interesting to analyze the pattern of memory accesses made by the
algorithms to compute an optimum BST, and optimize the running time of
each of the algorithms when run on the HMM model.

\subsection{BSTs optimum on both the RAM and the HMM}
When is the structure of the optimum BST the same on the HMM as on the
RAM model? In other words, is it possible to characterize when the
minimum-cost tree is the one that is optimum when the memory
configuration is uniform?

The following small example demonstrates that, in general, the
structure of an optimum tree on the uniform-cost RAM model can be very
different from the structure of an optimum tree on the HMM. To
discover this example, we used a computer program to perform an
exhaustive search.

Consider an instance of the problem of constructing an optimum BST on
the \HMM2 model, with $n=3$ keys.  The number of times $p_i$ that
the $i$-th key $x_i$ is accessed, for $1 \leq i \leq 3$, and the
number of times $q_j$ that the search argument lies between $x_j$
and $x_{j+1}$, for $0 \leq j \leq 3$, are:
\begin{align*}
p_i &= \seq{98, 72, 95} \\
q_j &= \seq{49, 20, 22, 84}
\end{align*}
The $p_i$'s and $q_j$'s are the frequencies of access. They are not
normalized to add up to $1$, but such a transformation could easily be
made without changing the optimum solution.

\begin{figure}[t]
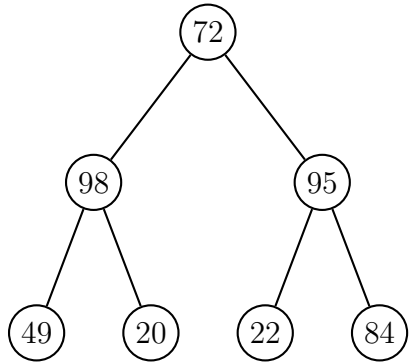

\begin{center}
\pstree{\Tcircle{$72$}} {\pstree{\Tcircle{$98$}} {\pstree{\Tcircle{$49$}} { }
\pstree{\Tcircle{$20$}} { }}
\pstree{\Tcircle{$95$}} {\pstree{\Tcircle{$22$}} { } \pstree{\Tcircle{$84$}} { }}}
\end{center}
\caption{An optimum BST on the unit-cost RAM model.}
\label{fig_conjsameRAM}
\hrule
\end{figure}

In this instance of the HMM model, there is one memory location each
whose cost is in $\{4$, $12$, $14$, $44$, $66$, $76$, $82\}$.
The optimum BST on the RAM model is shown in figure
\ref{fig_conjsameRAM}. Its cost on the RAM model with each location of
unit cost is $983$, while the cost of the same tree on this instance
of the HMM model is $16,752$.

\begin{figure}[t]
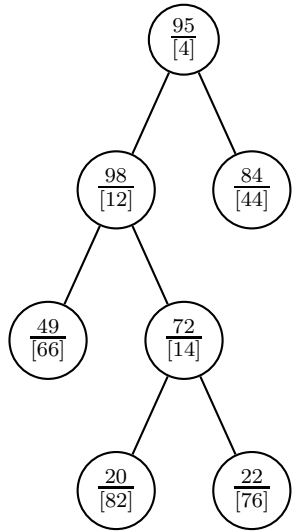

\begin{center}
\pstree{\Tcircle{$95 \over [4]$}} {\pstree{\Tcircle{$98 \over [12]$}}
{\pstree{\Tcircle{$49 \over [66]$}} { }
\pstree{\Tcircle{$72 \over [14]$}} {\pstree{\Tcircle{$20 \over [82]$}}
{ } \pstree{\Tcircle{$22 \over [76]$}} { }}}
\pstree{\Tcircle{$84 \over [44]$}} { }}
\end{center}
\caption{An optimum BST on the HMM model.}
\label{fig_conjsameHMM}
\hrule
\end{figure}

On the other hand, the BST over the same set of keys and frequencies
that is optimum on this instance of the HMM model is shown in
figure \ref{fig_conjsameHMM}. Its cost on the unit-cost RAM model is
$990$ and on the above instance of the HMM model is $16,730$. In figure
\ref{fig_conjsameHMM}, the nodes of the tree are labeled with the
frequency of the corresponding key, and the cost of the memory
location where the node is stored in square brackets.

\subsection{A monotonicity principle}
The dynamic programming algorithms, \textsc{algorithm Parts} of section
\ref{secn2hp2algorithm} and \textsc{algorithm TwoLevel} of section
\ref{secdynprog4hmm2}, iterate through the large number of possible ways
of partitioning the available memory locations between left and right
subtrees. It would be interesting to discover a monotonicity
principle, similar to the concave quadrangle inequality, which would
reduce the number of different options tried by the algorithms.

For the problem of constructing an optimum BST on the
\HMM2\ model with only two different memory costs, we were able to
\emph{disprove} the following conjectures by giving counter-examples:

\begin{conjecture}[Disproved]
\label{conj_mono1}
If $x_k$ is the root of an optimum subtree over the subset of keys
$x_i$ through $x_j$ in which $m$ cheap locations are assigned to the
left subtree, then the root of an optimum subtree over the same subset
of keys in which $m+1$ cheap locations are assigned to the left
subtree must have index no smaller than $k$.
\end{conjecture}

\noindent\textbf{Counter-example:} Consider an instance of the problem
of constructing an optimum BST on the \HMM2 model, with $n=7$ keys. In
this instance, there are $m_1=5$ cheap memory locations such that a
single access to a cheap location costs $c_1=5$, and $m_2=10$
expensive locations such that a single access to an expensive location
has cost $c_2=15$. The number of times $p_i$ that the $i$-th key
$x_i$ is accessed, for $1 \leq i \leq 7$, and the number of times
$q_j$ that the search argument lies between $x_j$ and $x_{j+1}$, for
$0 \leq j \leq 7$, are:
\begin{align*}
p_i &= \seq{2, 2, 2, 10, 4, 9, 5} \\
q_j &= \seq{6, 6, 7, 4, 1, 1, 9, 6}
\end{align*}
The $p_i$'s and $q_j$'s are the frequencies of access; they could
easily be normalized to add up to $1$.

An exhaustive search shows that the optimum BST with $n^{(L)}_1 = 0$
cheap locations assigned to the left subtree (and therefore, $4$ cheap
locations assigned to the right subtree), with total cost $1,890$, has
$x_3$ at the root.  The optimum BST with $n^{(L)}_1 = 1$ cheap
locations assigned to the left subtree (and $3$ cheap locations
assigned to the right subtree), with total cost $1,770$, has $x_2$ at
the root. This example disproves conjecture \ref{conj_mono1}.

\begin{conjecture}[Disproved]
\label{conj_mono2}
If $x_k$ is the root of an optimum subtree over the subset of keys
$x_i$ through $x_j$ in which $m$ cheap locations are assigned to the
left subtree, then in the optimum subtree over the same subset of keys
but with $x_{k+1}$ at the root, the left subtree must have assigned no
fewer than $m$ cheap locations.
\end{conjecture}

\noindent\textbf{Counter-example:}
Consider an instance of the problem again with $n=7$ keys. In
this instance, there are $m_1=5$ cheap memory locations such that a
single access to a cheap location costs $c_1=9$, and $m_2=10$
expensive locations such that a single access to an expensive location
has cost $c_2=27$. The number of times $p_i$ that the $i$-th key
$x_i$ is accessed, for $1 \leq i \leq 7$, and the number of times
$q_j$ that the search argument lies between $x_j$ and $x_{j+1}$, for
$0 \leq j \leq 7$, are:
\begin{align*}
p_i &= \seq{7, 3, 9, 3, 3, 6, 3} \\
q_j &= \seq{4, 9, 4, 5, 5, 7, 5, 9}
\end{align*}

As a result of an exhaustive search, we see that the optimum BST with
$x_4$ at the root, with total cost $3,969$, has $3$ cheap locations
assigned to the left subtree, and $1$ cheap location assigned to the
right subtree. However, the optimum BST with $x_5$ at the root, with
total cost $4,068$, has only $2$ cheap locations assigned to the left
subtree, and $2$ cheap locations assigned to the right subtree. This
example disproves conjecture \ref{conj_mono2}.

\begin{conjecture}[Disproved]
\label{conj_unimodal}
[\textbf{Conjecture of unimodality}]
The cost of an optimum BST with a fixed root $x_k$ is a unimodal
function of the number of cheap locations assigned to the left subtree.
\end{conjecture}

Conjecture \ref{conj_unimodal} would imply that we could substantially
improve the running time of \textsc{algorithm Parts} of section
\ref{secn2hp2algorithm}. The $h-1$ innermost loops of
\textsc{algorithm Parts} each perform a linear search for the optimum way
to partition the available memory locations from each level between
the left and right subtrees. If the conjecture were true, we could
perform a discrete unimodal search instead and reduce the overall
running time to $O((\log n)^{h-1} \cdot n^3)$.

\begin{figure}[t]
\begin{center}
\includegraphics[width=\linewidth]{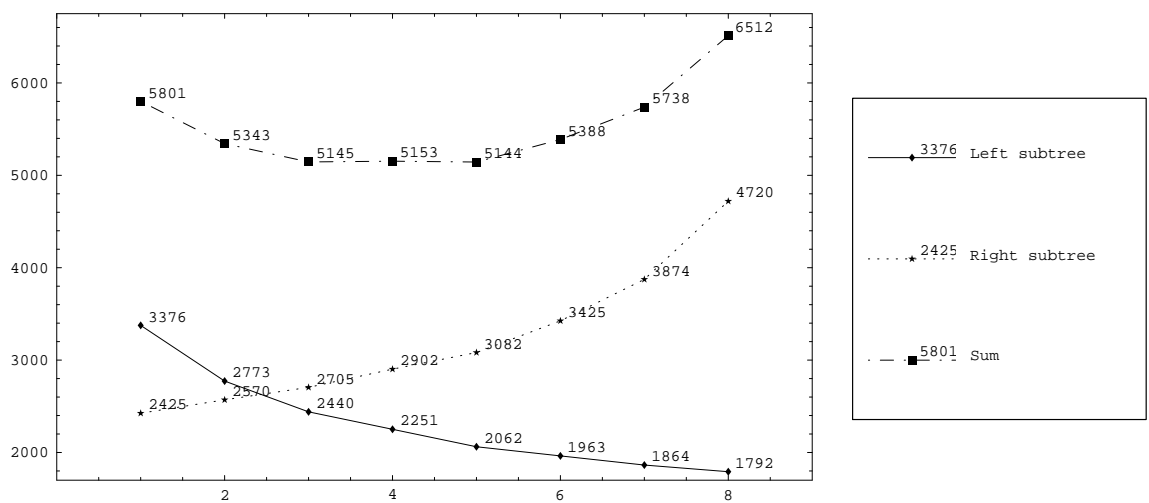}
\end{center}
\caption{The cost of an optimum BST is not a unimodal function.}
\label{graph_conjunimodal}
\hrule
\end{figure}

\noindent\textbf{Counter-example:}
A counter-example to conjecture \ref{conj_unimodal} is the binary
search tree over $n=15$ keys, where the frequencies of access are:
\begin{align*}
p_i &= \seq{2, 2, 9, 2, 1, 4, 10, 9, 9, 7, 5, 6, 9, 8, 10} \\
q_j &= \seq{1, 8, 8, 1, 3, 4, 6, 6, 6, 3, 3, 10, 8, 3, 4, 3}
\end{align*}
The instance of the HMM model has $m_1=7$ cheap memory locations of
cost $c_1=7$ and $m_2=24$ expensive locations of cost
$c_2=16$. Through an exhaustive search, we determined that the cost of
an optimum binary search tree with $x_8$ at the root exhibits the
behavior shown in the graph of figure \ref{graph_conjunimodal} as the
number $n^{(L)}_1$ of cheap locations assigned to the left subtree
varies from $0$ through $6$. (As the root, $x_8$ is always assigned to a
cheap location.) The graph of figure
\ref{graph_conjunimodal} plots the costs of the optimum left and right
subtrees of the root and their sum, as the number of cheap
locations assigned to the left subtree increases, or equivalently, as
the number of cheap locations assigned to the right subtree
decreases. (Note that the total cost of the BST is only a constant
more than the sum of the costs of the left and right subtrees since
the root is fixed.) We see from the graph that the cost of an optimum
BST with $n^{(L)}_1=4$ is greater than that for $n^{(L)}_1=3$ and
$n^{(L)}_1=5$; thus, the cost is not a unimodal function of
$n^{(L)}_1$.

\subsection{Dependence on the parameter $h$}
Downey and Fellows \cite{DowneyFellows} define a class of
parameterized problems, called \emph{fixed-parameter tractable}
(FPT).
\begin{definition}[Downey, Fellows \cite{DowneyFellows}]
A parameterized problem $L \subseteq \Sigma^* \times \Sigma^*$ is
\emph{fixed-parameter tractable} if there is an algorithm that correctly
decides for input $(x,y) \in \Sigma^* \times \Sigma^*$, whether $(x,y)
\in L$ in time $f(k) n^{\alpha}$, where $n$ is the size of the main
part of the input $x$, $\abs{x} = n$, $k$ is the integer parameter
which is the length of $y$, $\abs{y} = k$, $\alpha$ is a constant
independent of $k$, and $f$ is an arbitrary function.
\end{definition}

The best algorithm we have for the general problem, i.e., for
arbitrary $h$, is \textsc{algorithm Parts} of section
\ref{secn2hp2algorithm}, which runs in time $O(n^{h+2})$. Consider the
case where all $h$ levels in the memory hierarchy have roughly the
same number of locations, i.e., $m_1 = m_2 = \ldots = m_{h-1} =
\floor{n/h}$ and $m_h = \ceil{n/h}$. If the number of levels $h$ is a
parameter to the problem, it remains open whether this problem is
(strongly uniformly) fixed-parameter tractable---is there an algorithm
to construct an optimum BST that runs in time $O(f(h) n^{\alpha})$
where $\alpha$ is a constant independent of both $h$ and $n$? For
instance, is there an algorithm with running time $O(2^h n^{\alpha})$?
Recall that we have a top-down algorithm (\textsc{algorithm Split} of
section \ref{sec2nalgorithm}) that runs in time $O(2^n)$ for the case
$h=n$. A positive answer to this question would imply that it is
feasible to construct optimum BSTs over a large set of keys for a
larger range of values of $h$, in particular, even when $h = O(\log
n)$.


\renewcommand\bibname{References}
\bibliographystyle{myalpha}
\bibliography{msthesis}


\end{document}